\documentclass{amsart}%
\usepackage{amsfonts}
\usepackage{amsmath}
\usepackage{amssymb}
\usepackage{graphicx}%
\newtheorem{theorem}{Theorem}

\newtheorem{conjecture}[theorem]{Conjecture}

\newtheorem{definition}[theorem]{Definition}

\newtheorem{lemma}[theorem]{Lemma}

\newtheorem{proposition}[theorem]{Proposition}
\newtheorem{remark}[theorem]{Remark}

\begin{document}

\title[Holomorphic Sobolev Spaces]{Holomorphic Sobolev Spaces and the Generalized Segal--Bargmann Transform}

\author{Brian C. Hall}
\address[Brian C. Hall]{Department of Mathematics, University of Notre Dame, Notre Dame, Indiana 46556-4618, USA}
\email{bhall@nd.edu}
\thanks{The research of the first author was supported in part by the NSF grant DMS-0200649}

\author{Wicharn Lewkeeratiyutkul}
\address[Wicharn Lewkeeratiyutkul]{Department of Mathematics, Faculty of Science, Chulalongkorn University,
Bangkok 10330, Thailand}
\email{Wicharn.L@chula.ac.th}

\begin{abstract}
We consider the generalized Segal--Bargmann transform $C_{t}$ for a compact
group $K,$ introduced in B. C. Hall, \textit{J. Funct. Anal.} \textbf{122}
(1994), 103-151. Let $K_{\mathbb{C}}$ denote the complexification of $K.$ We
give a necessary-and-sufficient pointwise growth condition for a holomorphic
function on $K_{\mathbb{C}}$ to be in the image under $C_{t}$ of $C^{\infty
}(K).$ We also characterize the image under $C_{t}$ of Sobolev spaces on $K.$
The proofs make use of a holomorphic version of the Sobolev embedding theorem.
\end{abstract}

\maketitle

\tableofcontents

\section{Introduction and statement of results\label{intro.sec}}

The Segal--Bargmann transform, in a form convenient for the purposes of this
paper, is the map $C_{t}:L^{2}(\mathbb{R}^{d})\rightarrow\mathcal{H}%
(\mathbb{C}^{d})$ given by%
\begin{equation}
C_{t}f(z)=\int_{\mathbb{R}^{d}}(2\pi t)^{-d/2}e^{-(z-x)^{2}/2t}f(x)\,dx,\quad
z\in\mathbb{C}^{d}. \label{ctdef1}%
\end{equation}
Here $(z-x)^{2}=(z_{1}-x_{1})^{2}+\cdots+(z_{d}-x_{d})^{2}$ and $\mathcal{H}%
(\mathbb{C}^{d})$ denotes the space of (entire) holomorphic functions on
$\mathbb{C}^{d}.$ It is easily verified that the integral in (\ref{ctdef1}) is
absolutely convergent for all $z\in\mathbb{C}^{d}$ and that the result is a
holomorphic function of $z.$ If we restrict attention to $z\in\mathbb{R}^{d},$
then we may recognize the function%
\begin{equation}
(2\pi t)^{-d/2}e^{-(z-x)^{2}/2t} \label{heat.kernel}%
\end{equation}
as the \textit{heat kernel} for $\mathbb{R}^{d},$ that is, the integral kernel
for the time-$t$ heat operator. This means that $C_{t}f$ may alternatively be
described as
\begin{equation}
C_{t}f=\text{ analytic continuation of }e^{t\Delta/2}f. \label{ctdef2}%
\end{equation}
Here the analytic continuation is from $\mathbb{R}^{d}$ to $\mathbb{C}^{d}$
with $t$ fixed, and $e^{t\Delta/2}$ is the time-$t$ (forward) heat operator.
(We take the Laplacian to be a negative operator and follow the probabilists'
normalization of the heat operator.)

\begin{theorem}[Segal--Bargmann]\label{sb.thm1}For each $t>0,$ the map $C_{t}$ is a unitary
map of $L^{2}(\mathbb{R}^{d})$ onto $\mathcal{H}L^{2}(\mathbb{C}^{d},\nu
_{t}).$ Here $\mathcal{H}L^{2}(\mathbb{C}^{d},\nu_{t})$ denotes the space of
holomorphic functions that are square-integrable with respect to the measure
$\nu_{t}(z)\,dz,$ where $dz$ denotes $2d$-dimensional Lebesgue measure on
$\mathbb{C}^{d}$ and where $\nu_{t}$ is the density given by%
\[
\nu_{t}(x+iy)=(\pi t)^{-d/2}e^{-y^{2}/t},\quad x,y\in\mathbb{R}^{d}.
\]
\end{theorem}

Our normalization of the Segal--Bargmann transform is different from that of
Segal \cite{Se} and Bargmann \cite{Ba1}; see \cite{mexnotes}, \cite{lpbounds},
or \cite{H1} for a comparison of normalizations. Note that the function
$\nu_{t}$ is simply the heat kernel at the origin in the $y$ variable,
evaluated at time $t/2.$ (That is, to get $\nu_{t},$ put $z=0$ in
(\ref{heat.kernel}), replace $x$ with $y$, and replace $t$ by $t/2.$)

One of the distinctive features of $L^{2}$ spaces of holomorphic functions is
that \textquotedblleft pointwise evaluation is continuous.\textquotedblright%
\ This means, in the present setting, that for each $z\in\mathbb{C}^{d},$ the
map $F\rightarrow F(z)$ is a continuous linear functional on $\mathcal{H}%
L^{2}(\mathbb{C}^{d},\nu_{t}).$ One can shown (adapting a result of Bargmann
\cite{Ba1} to our normalization) that the norm of the \textquotedblleft
evaluation at $z$\textquotedblright\ functional is precisely $(4\pi
t)^{-d/4}e^{y^{2}/2t}.$ This means that elements $F$ of $\mathcal{H}%
L^{2}(\mathbb{C}^{d},\nu_{t})$ satisfy the pointwise bounds%
\begin{equation}
|F(x+iy)|^{2}\leq Ae^{y^{2}/t},\label{rd.bound1}%
\end{equation}
where the optimal value of $A$ is $(4\pi t)^{-d/2}\left\Vert F\right\Vert
_{L^{2}(\mathbb{C}^{d},\nu_{t})}^{2}$. Conversely, if a holomorphic function
$F$ satisfies a polynomially better bound, say,%
\begin{equation}
|F(x+iy)|^{2}\leq A\frac{e^{y^{2}/t}}{1+(x^{2}+y^{2})^{d+\varepsilon}}%
,\quad\varepsilon>0,\label{rd.bound2}%
\end{equation}
then, by direct calculation, $F$ will be square-integrable with respect to the
measure $\nu_{t}(z)\,dz$ and thus will be in $\mathcal{H}L^{2}(\mathbb{C}%
^{d},\nu_{t}).$

Theorem \ref{sb.thm1} characterizes the image under $C_{t}$ of $L^{2}%
(\mathbb{R}^{d})$ exactly as a holomorphic $L^{2}$ space over $\mathbb{C}^{d}.
$ The image of $L^{2}(\mathbb{R}^{d})$ can also be characterized by the
necessary pointwise bounds (\ref{rd.bound1}) and the slightly stronger
sufficient pointwise bounds (\ref{rd.bound2}). It is natural to ask in
addition for a characterization of other spaces of functions, for example, the
Schwarz space. The \textquotedblleft polynomial closeness\textquotedblright%
\ between the necessary bounds (\ref{rd.bound1}) and the sufficient bounds
(\ref{rd.bound2}) is a key ingredient in the following result of Bargmann
\cite[Theorem 1.7]{Ba2} (adapted, as always, to our normalization of the transform).

\begin{theorem}[Bargmann]\label{bargmann.thm}Let $\mathcal{S}(\mathbb{R}^{d})$ denote the
Schwarz space. If $F$ is a holomorphic function on $\mathbb{C}^{d}$ then there
exists $f\in\mathcal{S}(\mathbb{R}^{d})$ with $C_{t}f=F$ if and only if $F$
satisfies%
\[
|F(x+iy)|^{2}\leq A_{n}\frac{e^{y^{2}/t}}{[1+(x^{2}+y^{2})]^{2n}}
\]
for some sequence of constants $A_{n},$ $n=1,2,3,\ldots.$
\end{theorem}

See also \cite{lpbounds} for related results. Roughly speaking, smoothness of
$f$ gives a polynomial improvement in the behavior of $F$ (compared to
(\ref{rd.bound1})) in the imaginary ($y$) directions, while decay at infinity
of $f$ gives polynomial improvement of $F$ in the real ($x$) directions.

The purpose of this paper is to obtain similar results for the generalized
Segal--Bargmann transform introduced in \cite{H1}. (The paper \cite{H1}
was motivated by results of Gross \cite{Gr}. For more information
about the generalized Segal--Bargmann transform and its connections to the
work of Gross, see \cite{harmonic} and \cite{ergodic}.) Let $K$ be an arbitrary
connected compact Lie group. Fix once and for all a bi-invariant Riemannian
metric on $K$ and let $\Delta_{K}$ denote the (negative) Laplacian operator
with respect to this metric. Let $K_{\mathbb{C}}$ denote the complexification
of $K,$ which is a certain complex Lie group containing $K$ as a maximal
compact subgroup. (For example, if $K=\mathsf{U}(n)$ then $K_{\mathbb{C}%
}=\mathsf{GL}(n;\mathbb{C}).$) Let $dx$ denote the Haar measure on $K,$
normalized to coincide with the Riemannian volume measure. Then, by analogy to
the $\mathbb{R}^{d}$ case, we define a map $C_{t}:L^{2}(K,dx)\rightarrow
\mathcal{H}(K_{\mathbb{C}})$ by%
\[
C_{t}f=\text{analytic continuation of }e^{t\Delta_{K}/2}f.
\]
It can be shown \cite[Sect. 4]{H1} that for any $f\in L^{2}(K,dx)$ and any
fixed $t>0,$ $e^{t\Delta_{K}/2}f$ admits a unique analytic continuation from
$K$ to $K_{\mathbb{C}}.$ One of the main results of \cite{H1} is the following.

\begin{theorem}
\label{sb.thm2}For each $t>0$ there exists a smooth positive function $\nu
_{t}$ on $K_{\mathbb{C}}$ such that $C_{t}$ is a unitary isomorphism of
$L^{2}(K,dx)$ onto $\mathcal{H}L^{2}(K_{\mathbb{C}},\nu_{t}).$ Here
$\mathcal{H}L^{2}(K_{\mathbb{C}},\nu_{t})$ denotes the space of holomorphic
functions on $K_{\mathbb{C}}$ that are square-integrable with respect to the
measure $\nu_{t}(g)\,dg,$ where $dg$ is the Haar measure on $K_{\mathbb{C}}. $
\end{theorem}

We will use a convenient normalization of the Haar measure on $K_{\mathbb{C}%
},$ given in (\ref{dg.form}) below. As in the $\mathbb{R}^{d}$ case, the
function $\nu_{t}$ is the ``heat kernel at the origin in the imaginary
variables.'' This means, more precisely, that $\nu_{t}$ is the heat kernel at
the identity coset for the noncompact symmetric space $K_{\mathbb{C}}/K,$
viewed as a bi-$K$-invariant function on $K_{\mathbb{C}}$ (and evaluated at
time $t/2$). There is an explicit formula for $\nu_{t},$ due to Gangolli,
which we will make use of repeatedly in what follows. (See (\ref{nut.function}).)

In this paper, we consider the image under $C_{t}$ of spaces other than
$L^{2}(K).$ Since $K$ is compact, there is no behavior at infinity to worry
about, and therefore the natural function spaces to consider are ones with
various degrees of smoothness. We will consider Sobolev spaces on $K$ and also
$C^{\infty}(K).$ In particular we will give (Theorem \ref{main.thm}) a single
necessary-and-sufficient pointwise condition that a holomorphic function must
satisfy in order to be in the image under $C_{t}$ of $C^{\infty}(K).$ This
result is the analog for a compact group of Bargmann's result (Theorem
\ref{bargmann.thm}) for $\mathbb{R}^{d}.$

To describe our results we introduce \textit{polar coordinates} on
$K_{\mathbb{C}},$ which are analogous to the coordinates $z=x+iy$ on
$\mathbb{C}^{d}.$ If $\mathfrak{k}$ denotes the Lie algebra of $K,$ then the
Lie algebra of $K_{\mathbb{C}}$ is $\mathfrak{k}_{\mathbb{C}}:=\mathfrak{k}%
+i\mathfrak{k},$ and so we may consider the exponential mapping from
$\mathfrak{k}+i\mathfrak{k}$ into $K_{\mathbb{C}}.$

\begin{proposition}[Polar Coordinates]\label{polar.prop}For each $g$ in $K_{\mathbb{C}}$ there
exists a unique $x$ in $K$ and $Y$ in $\mathfrak{k}$ such that%
\begin{equation}
g=xe^{iY},\quad x\in K,\,Y\in\mathfrak{k}. \label{polar}%
\end{equation}
Furthermore, the map $(x,Y)\rightarrow xe^{iY}$ is a diffeomorphism of
$K\times\mathfrak{k}$ with $K_{\mathbb{C}}.$
\end{proposition}

This is a standard result in the case in which $K$ is semisimple \cite[Theorem
6.31]{Kn} and is easily extended to the general case, as discussed in Section
11 of \cite{H1}. Consider, for example, the case $K=\mathsf{U}(n). $ Then
$K_{\mathbb{C}}=\mathsf{GL}(n;\mathbb{C})$ and the elements of $\mathfrak{k}%
=\mathsf{u}(n)$ are skew-self-adjoint matrices. Thus for $Y\in\mathfrak{k},$
$iY$ will be self-adjoint and $e^{iY}$ will be self-adjoint and positive. Thus
the decomposition $g=xe^{iY}$ for a matrix $g\in\mathsf{GL}(n;\mathbb{C})$ is
the ordinary polar decomposition into the product of a unitary matrix $x$ and
a positive matrix $e^{iY}.$

Now let $\Phi$ be the unique Ad-$K$-invariant function on $\mathfrak{k}$ whose
restriction to any maximal commutative subspace $\mathfrak{t}$ of
$\mathfrak{k}$ is given by%
\begin{equation}
\Phi(H)=\prod_{\alpha\in R^{+}}\frac{\alpha(H)}{\sinh\alpha(H)},\quad
H\in\mathfrak{t}. \label{phi.def}%
\end{equation}
Here $R\subset\mathfrak{t}^{\ast}$ denotes the set of real roots of
$\mathfrak{k}$ relative to $\mathfrak{t}$ and $R^{+}$ denotes a set of
positive roots for this root system. We are now ready to state the main result
of this paper.

\begin{theorem}
\label{main.thm}Suppose $F$ is a holomorphic function on $K_{\mathbb{C}}$ and
$t$ is a fixed positive number. Then there exists $f\in C^{\infty}(K)$ with
$F=C_{t}f$ if and only if $F$ satisfies%
\begin{equation}
\left\vert F(xe^{iY})\right\vert ^{2}\leq A_{n}\Phi(Y)\frac{e^{|Y|^{2}/t}%
}{(1+|Y|^{2})^{2n}} \label{main.form}%
\end{equation}
for some sequence of constants $A_{n},$ $n=1,2,3,\ldots.$
\end{theorem}

In the right-hand side of (\ref{main.form}) we think of $\mathfrak{k}$ as the
tangent space to $K$ at the identity. Then $\left\vert Y\right\vert $ is
computed with respect to the restriction of the bi-invariant metric on $K$ to
$\mathfrak{k}=T_{e}(K).$

Since the function $\Phi$ plays a critical role in this paper, it is worth
taking a moment to consider its behavior. If $K$ is commutative then there are
no roots and so we have
\[
\Phi(Y)\equiv1\quad\text{(commutative case).}%
\]
If $K$ is semisimple then the roots span $\mathfrak{t}^{\ast}$ and as a result
the function $\Phi$ has exponential decay at infinity. For example, consider
the rank-one case $K=\mathsf{SU}(2)$ and equip $\mathsf{SU}(2)$ with the
bi-invariant Riemannian metric whose restriction to $\mathsf{su}%
(2)=T_{e}(\mathsf{SU}(2))$ is given by $\left\vert Y\right\vert ^{2}%
=2~\mathrm{trace}(Y^{\ast}Y).$ Then we have%
\[
\Phi(Y)=\frac{\left\vert Y\right\vert }{\sinh|Y|}\quad\text{(}\mathsf{SU}%
(2)\text{ case)}%
\]
for all $Y$ in $\mathsf{su}(2).$ The function $\Phi$ is related to the
exponential growth (in the noncommutative case) of Haar measure on
$K_{\mathbb{C}}$. Specifically, the Haar measure on $K_{\mathbb{C}}$ can be
written in polar coordinates as follows:%
\begin{equation}
dg=\frac{1}{\Phi(Y)^{2}}\,dx\,dY, \label{dg.form}%
\end{equation}
where $dx$ is the Haar measure on $K$ and $dY$ is the Lebesgue measure on
$\mathfrak{k}$ (normalized by the inner product). (See \cite[Lem. 5]{H3}.)

To understand the significance of the bounds (\ref{main.form}), we need to
look at the expression for the measure $\nu_{t}(g)\,dg.$ The \textit{function}
$\nu_{t}$ has the following expression, due to Gangolli \cite[Prop. 3.2]{Ga}:%
\begin{equation}
\nu_{t}(xe^{iY})=c_{t}\Phi(Y)e^{-|Y|^{2}/t},\quad x\in K,\,Y\in\mathfrak{k},
\label{nut.function}%
\end{equation}
where%
\begin{equation}
c_{t}=(\pi t)^{-d/2}e^{-\left\vert \delta\right\vert ^{2}t}. \label{ct.def}%
\end{equation}
In (\ref{ct.def}), $\delta$ is half the sum of the positive roots for $K$ and
$d=\dim K.$ (See also \cite[Eq. (11)]{H3}.) By combining (\ref{nut.function})
and (\ref{dg.form}) we obtain an expression for the \textit{measure} $\nu
_{t}(g)\,dg,$ namely,%
\begin{equation}
\nu_{t}(g)\,dg=c_{t}\frac{e^{-|Y|^{2}/t}}{\Phi(Y)}\,dx\,dY,\quad g=xe^{iY}.
\label{nut.measure}%
\end{equation}

Let us compare to the $\mathbb{R}^{d}$ case, where $\nu_{t}(x+iy)=(\pi
t)^{-d/2}e^{-|y|^{2}/t}.$ If $K$ is commutative then $\Phi$ is identically
equal to one and things behave as in the $\mathbb{R}^{d}$ case. If $K$ is
semisimple, then $\Phi(Y)$ decays exponentially. This means that although the
function $\nu_{t}$ has \textit{faster} decay at infinity than $e^{-|Y|^{2}%
/t},$ the measure $\nu_{t}(g)\,dg$ (which is the quantity that really matters)
has \textit{slower} decay at infinity than the measure $e^{-|Y|^{2}%
/2}\,dx\,dY.$ The slower decay at infinity of the heat kernel measure reflects
that (if $K$ is semisimple) $K_{\mathbb{C}}/K$ has a lot of negative
curvature. The negative curvature causes the heat to flow out to infinity
faster than in the Euclidean case, which makes the heat kernel measure larger
near infinity than in the Euclidean case.

The paper \cite{H3} establishes the following pointwise bounds for elements
$F$ of $\mathcal{H}L^{2}(K_{\mathbb{C}},\nu_{t})$%
\begin{equation}
\left\vert F(xe^{iY})\right\vert ^{2}\leq A\Phi(Y)e^{|Y|^{2}/t}.
\label{kc.bounds}%
\end{equation}
In the semisimple case, this bound is better (as a function of $Y$) than in
the $\mathbb{R}^{d}$ case, because of the exponentially decaying factor
$\Phi(Y).$ Intuitively, the reason for this is the slower decay at infinity of
the measure (\ref{nut.measure}): If $F$ is to be square-integrable with
respect to the slower-decaying measure, then $F$ must have correspondingly
better behavior at infinity. (Nevertheless, actually proving the bounds in
(\ref{kc.bounds}) is not especially easy; see Section \ref{conclude.sec}.) As
in the $\mathbb{R}^{d}$ case, we can see by direct calculation (using
(\ref{nut.measure})) that if a holomorphic function $F$ on $K_{\mathbb{C}}$
satisfies polynomially stronger bounds than (\ref{kc.bounds}), say,%
\begin{equation}
\left\vert F(xe^{iY})\right\vert ^{2}\leq A\Phi(Y)\frac{e^{|Y|^{2}/t}%
}{(1+\left\vert Y\right\vert ^{2})^{d/2+\varepsilon}},\quad\varepsilon>0,
\label{kc.suffbds}%
\end{equation}
then $F$ is square-integrable with respect to the measure in
(\ref{nut.measure}) and is therefore in $\mathcal{H}L^{2}(K_{\mathbb{C}}%
,\nu_{t}).$

It is the polynomial closeness of the necessary bounds (\ref{kc.bounds}) and
the sufficient bounds (\ref{kc.suffbds}) that is the key to the proof of
Theorem \ref{main.thm}. Specifically, instead of considering $\mathcal{H}%
L^{2}(K_{\mathbb{C}},\nu_{t}),$ which is the image under $C_{t}$ of
$L^{2}(K),$ we will consider the image under $C_{t}$ of the Sobolev space
$H^{2n}(K),$ consisting of functions on $K$ having all derivatives up to order
$2n$ in $L^{2}.$ We will give necessary pointwise bounds and sufficient
pointwise bounds for the image of $H^{2n}(K).$ These bounds are the same as
(\ref{kc.bounds}) and (\ref{kc.suffbds}), except with an extra factor of
$(1+\left\vert Y\right\vert ^{2})^{-2n}$ on the right-hand side. (Compare
(\ref{kc.bounds}) and (\ref{kc.suffbds}) to (\ref{hsob.est}) and
(\ref{sobolev.est}).) The polynomial closeness of the two sets of bounds means
that the necessary bounds for one value of $n$ become sufficient for some
slightly smaller value of $n$. Thus, after intersecting over all $n$ we obtain
the single necessary-and-sufficient condition on the image of $C^{\infty}(K)$
given in Theorem \ref{main.thm}.

In the remainder of this section, we state the results concerning Sobolev
spaces and describe the strategy for proving Theorem \ref{main.thm}. The
Sobolev space $H^{2n}(K)$ can also be described as the set of $f$ in
$L^{2}(K)$ such that $\Delta_{K}^{n}f$ (computed in the distributional sense)
is again in $L^{2}(K).$ We then think of $C^{\infty}(K)$ as the intersection
of $H^{2n}(K)$ over $n=1,2,3,\ldots.$ In Section \ref{embed.sec}, we will give
the following (easy) characterization of the image of $H^{2n}(K)$ under
$C_{t}.$

\begin{theorem}
\label{sobolev.thm1}For $f\in L^{2}(K),$ let $F=C_{t}f.$ Then $f$ is in
$H^{2n}(K)$ if and only if $\Delta_{K}^{n}F\in\mathcal{H}L^{2}(K_{\mathbb{C}%
},\nu_{t}).$
\end{theorem}

In computing $\Delta_{K}^{n}F,$ we regard $\Delta_{K}$ as a left-invariant
differential operator on $K_{\mathbb{C}}$; see Section \ref{embed.sec} for
details. We wish to think of the set of holomorphic functions $F$ with
$F\in\mathcal{H}L^{2}(K_{\mathbb{C}},\nu_{t})$ and $\Delta_{K}^{n}%
F\in\mathcal{H}L^{2}(K_{\mathbb{C}},\nu_{t})$ as a sort of \textquotedblleft
holomorphic Sobolev space.\textquotedblright\ Now, it may seem odd at first to
speak of Sobolev spaces in the holomorphic context. After all, every element
$F$ of $\mathcal{H}L^{2}(K_{\mathbb{C}},\nu_{t})$ is automatically infinitely
differentiable, and $\Delta_{K}^{n}F$ is automatically holomorphic again.
Nevertheless, given $F\in\mathcal{H}L^{2}(K_{\mathbb{C}},\nu_{t}),$ there is
no reason that $\Delta_{K}^{n}F$ must be again square-integrable with respect
to $\nu_{t}(g)\,dg.$ Thus having $\Delta_{K}^{n}F$ be in $\mathcal{H}%
L^{2}(K_{\mathbb{C}},\nu_{t})$ is a nontrivial \textquotedblleft
regularity\textquotedblright\ condition on $F.$

\begin{definition}
\label{sobolev.def}The $2n^{\text{th}}$ \textbf{holomorphic Sobolev space} on
$K_{\mathbb{C}},$ denoted $\mathcal{H}^{2n}(K_{\mathbb{C}},\nu_{t}),$ is the
space of holomorphic functions $F$ on $K_{\mathbb{C}}$ such that $F\in
L^{2}(K_{\mathbb{C}},\nu_{t})$ and $\Delta_{K}^{n}F\in L^{2}(K_{\mathbb{C}%
},\nu_{t}).$
\end{definition}

Note that on $K,$ membership in the Sobolev space $H^{2n}(K)$ is a local
condition: Since $K$ is compact, if $f$ is in $H^{2n}(K)$ locally then it is
in $H^{2n}(K)$ globally. By contrast, membership in the holomorphic Sobolev
space is a condition on the behavior of the function at infinity: If $F$ is
holomorphic then $\Delta_{K}^{n}F$ is automatically square-integrable locally,
and it is only the behavior at infinity that one needs to worry about. Using
holomorphic Fourier series (see \cite[Sect. 8]{H1}) it is easy to show that if
$F$ is \textit{any} holomorphic function on $K_{\mathbb{C}}$ such that
$\Delta_{K}^{n}F\in\mathcal{H}L^{2}(K_{\mathbb{C}},\nu_{t}),$ $F $ itself will
automatically be in $L^{2}(K_{\mathbb{C}},\nu_{t})$ and therefore $F$ will be
in $\mathcal{H}^{2n}(K_{\mathbb{C}},\nu_{t}).$ The same sort of reasoning
shows that if $F\in\mathcal{H}^{2n}(K_{\mathbb{C}},\nu_{t}),$ then $\Delta
_{K}^{m}F\in\mathcal{H}L^{2}(K_{\mathbb{C}},\nu_{t})$ for $1\leq m\leq n$ and,
more generally, all left-invariant derivatives of $F$ up to order $2n$ are in
$\mathcal{H}L^{2}(K_{\mathbb{C}},\nu_{t}).$

We wish to relate membership of a function $F$ in the holomorphic Sobolev
space to behavior of $F$ at infinity. Such a relationship can be thought of as
a holomorphic version of the Sobolev embedding theorem: existence of
derivatives in $L^{2}$ translates into improved pointwise behavior of the
function itself. Such a result is obtained in Section \ref{embed.sec} by
estimating the reproducing kernel for the $2n^{\text{th}}$ holomorphic Sobolev
space, leading to the following.

\begin{theorem}
[Holomorphic Sobolev embedding theorem]\label{embed.thm}If $F$ belongs to the
$2n^{\text{th}}$ holomorphic Sobolev space $\mathcal{H}^{2n}(K_{\mathbb{C}%
},\nu_{t}),$ then for some constant $A$ (depending on $F$ and $n$) we have%
\begin{equation}
\left\vert F(xe^{iY})\right\vert ^{2}\leq A\frac{\Phi(Y)e^{|Y|^{2}/t}%
}{(1+\left\vert Y\right\vert ^{2})^{2n}}, \label{hsob.est}%
\end{equation}
where $\Phi$ is as in (\ref{phi.def}).
\end{theorem}

This bound is the same as in (\ref{kc.bounds}) except for the extra factor of
$(1+\left|  Y\right|  ^{2})^{2n}$ in the denominator.

In this holomorphic setting we can also reason in the opposite direction, in a
way that is impossible in ordinary Sobolev spaces. That is, good pointwise
bounds imply membership in holomorphic Sobolev spaces. In Section
\ref{toeplitz.sec}, we will prove the following result of this sort.

\begin{theorem}
\label{sobolev.thm2}If $F$ is a holomorphic function on $K_{\mathbb{C}}$ then
$F$ belongs to the $2n^{\text{th}}$ holomorphic Sobolev space $\mathcal{H}%
^{2n}(K_{\mathbb{C}},\nu_{t})$ if and only if%
\begin{equation}
\int_{K_{\mathbb{C}}}\left|  F(g)\right|  ^{2}(1+\left|  Y\right|  ^{2}%
)^{2n}\nu_{t}(g)\,dg<\infty,\quad g=xe^{iY}. \label{sobolev.int}%
\end{equation}
That is, $\mathcal{H}^{2n}(K_{\mathbb{C}},\nu_{t})=\mathcal{H}L^{2}%
(K_{\mathbb{C}},(1+\left|  Y\right|  ^{2})^{2n}\nu_{t}(g)\,dg).$
\end{theorem}

This result is proved by means of integration by parts. Specifically, the
function $(1+\left\vert Y\right\vert ^{2})^{2n}$ has the same behavior as a
certain logarithmic-type derivative of $\nu_{t},$ and it is this
logarithmic-type derivative that comes out of the integration by parts.

Suppose, now, that $F$ is a holomorphic function satisfying polynomially
better bounds than those in Theorem \ref{embed.thm}, say,%
\begin{equation}
\left\vert F(xe^{iY})\right\vert ^{2}\leq A\Phi(Y)\frac{e^{|Y|^{2}/t}%
}{(1+\left\vert Y\right\vert ^{2})^{2n+d/2+\varepsilon}}\quad\text{(}d=\dim
K\text{).} \label{sobolev.est}%
\end{equation}
Then Theorem \ref{sobolev.thm2}, together with the expression
(\ref{nut.measure}) for the measure $\nu_{t}(g)\,dg,$ shows that $F$ belongs
to $\mathcal{H}^{2n}(K_{\mathbb{C}},\nu_{t}).$

By combining Theorems \ref{embed.thm} and \ref{sobolev.thm2} we can
immediately obtain Theorem \ref{main.thm}. Consider $F\in\mathcal{H}%
L^{2}(K_{\mathbb{C}},\nu_{t})$ and let $f=C_{t}^{-1}F.$ In one direction, if
$F$ satisfies the bounds in (\ref{main.form}) for a given $n,$ then
(\ref{sobolev.est}) shows that the integral in (\ref{sobolev.int}) is finite
for all $n^{\prime}$ with $n^{\prime}<n-d/4$. Thus if (\ref{main.form}) holds
for all $n,$ then so does (\ref{sobolev.int}). In such cases, $F\in
\mathcal{H}^{2n}(K_{\mathbb{C}},\nu_{t})$ and $f\in H^{2n}(K)$ for all $n,$
which implies that $f\in C^{\infty}(K).$

In the other direction, if $f\in C^{\infty}(K),$ then certainly $f\in
H^{2n}(K)$ for all $n$ and, therefore, $F\in\mathcal{H}^{2n}(K_{\mathbb{C}%
},\nu_{t})$ for all $n.$ This, by Theorem \ref{embed.thm}, implies that $F$
satisfies the bounds in Theorem \ref{main.thm}.

\section{The holomorphic Sobolev embedding theorem\label{embed.sec}}

The goal of this section is to estimate the reproducing kernel for the
holomorphic Sobolev spaces introduced in Definition \ref{sobolev.def}, leading
to a proof of the holomorphic Sobolev embedding theorem (Theorem
\ref{embed.thm}). We begin with the proof of Theorem \ref{sobolev.thm1}, which
asserts that the image under $C_{t}$ of the Sobolev space $H^{2n}(K)$ is the
holomorphic Sobolev space $\mathcal{H}^{2n}(K_{\mathbb{C}},\nu_{t})$ described
in Definition \ref{sobolev.def}. This result holds essentially because both
the heat operator and analytic continuation commute with $\Delta_{K}.$

Before turning to the proof of the theorem, we explain how $\Delta_{K}$ is to
be viewed as a differential operator on $K_{\mathbb{C}}.$ On $K,$ we have
$\Delta_{K}=\sum_{k}X_{k}^{2},$ where $X_{1,}\ldots,X_{d}$ is an orthonormal
basis for $\mathfrak{k}$ and each $X_{k}$ is viewed as a left-invariant
differential operator on $K.$ Since $\mathfrak{k}\subset\mathfrak{k}%
_{\mathbb{C}},$ we may also regard each $X_{k}$ as a left-invariant
differential operator on $K_{\mathbb{C}}.$ Then on $K_{\mathbb{C}},$ we define
$\Delta_{K}=\sum_{k}X_{k}^{2}.$ This means that, for any $C^{\infty}$ function
$\phi$ on $K_{\mathbb{C}}$ we have%
\begin{equation}
(\Delta_{K}\phi)(g)=\sum_{k=1}^{d}\left.  \frac{d^{2}}{dt^{2}}\phi\left(
ge^{tX_{k}}\right)  \right\vert _{t=0}. \label{lap.kc}%
\end{equation}
Note that although $\Delta_{K}$ is a bi-$K$-invariant operator on $K,$ on
$K_{\mathbb{C}}$, $\Delta_{K}$ is only left-$K_{\mathbb{C}}$-invariant and not
bi-$K_{\mathbb{C}}$-invariant. (When applied to holomorphic functions,
however, $\Delta_{K}$ coincides with the analogously defined
right-$K_{\mathbb{C}}$-invariant operator.) The operator $\Delta_{K}$
preserves the space of holomorphic functions on $K_{\mathbb{C}}.$ Furthermore,
if a function $f$ on $K$ admits an analytic continuation to a holomorphic
function on $K_{\mathbb{C}},$ then $\Delta_{K}f$ also has an analytic
continuation to $K_{\mathbb{C}},$ given by $(\Delta_{K}f)_{\mathbb{C}}%
=\Delta_{K}(f_{\mathbb{C}}),$ where $(\cdot)_{\mathbb{C}}$ denotes analytic
continuation. That is, $\Delta_{K}$ commutes with analytic continuation. We
now proceed with the proof of Theorem \ref{sobolev.thm1}.

\begin{proof}
[Proof of Theorem \ref{sobolev.thm1}]On $K,$ we consider $\Delta_{K}$, defined
at first on the space of finite linear combinations of matrix entries. (A
matrix entry is a function on $K$ of the form $f(x)=\mathrm{trace}(\pi(x)A),$
where $\pi$ is an irreducible representation of $K$ acting on a
finite-dimensional space $V$ and where $A$ is a linear operator on $V.$) Then
$\Delta_{K}^{n}$ is essentially self-adjoint on this space. This holds because
$L^{2}(K)$ is the orthogonal direct sum of the spaces of matrix entries (as
the representation varies over equivalence classes of irreducible
representations of $K$) and the restriction of $\Delta_{K}^{n}$ to each space
of matrix entries (for a fixed representation) is a real multiple of the
identity. Then we define $H^{2n}(K)$ as the domain of the closure of the
operator \textquotedblleft$\Delta_{K}^{n}$ on finite linear combinations of
matrix entries.\textquotedblright\ It is not hard to show that $H^{2n}(K)$
coincides with the space of all $f\in L^{2}(K)$ such that $\Delta_{K}^{n}f$
(computed in the distribution sense) is in $L^{2}(K).$ Furthermore, if $f\in
H^{2n}(K),$ then $f$ is in the domain of $\Delta_{K}^{m}$ for all $m$ between
$1$ and $n.$

To compute the image under $C_{t}$ of $H^{2n}(K),$ suppose, first, that $f\in
H^{2n}(K).$ Since $\Delta_{K}$ commutes with the heat operator $e^{t\Delta
_{K}/2}$ and with analytic continuation, we have $C_{t}(\Delta_{K}%
^{n}f)=\Delta_{K}^{n}C_{t}f,$ which shows that $C_{t}f\in\mathcal{H}%
^{2n}(K_{\mathbb{C}},\nu_{t}).$

In the reverse direction, it suffices to show that $\Delta_{K}^{n}$ is a
symmetric operator on the domain $\mathcal{H}^{2n}(K_{\mathbb{C}},\nu
_{t})\subset\mathcal{H}L^{2}(K_{\mathbb{C}},\nu_{t}).$ After all, since
$C_{t}$ is unitary, $C_{t}\Delta_{K}^{n}C_{t}^{-1}$ is self-adjoint on
$C_{t}(H^{2n}(K)).$ But we have just shown that $C_{t}(H^{2n}(K))\subset
\mathcal{H}^{2n}(K_{\mathbb{C}},\nu_{t})$ and that $C_{t}\Delta_{K}^{n}%
C_{t}^{-1}F=\Delta_{K}^{n}F$ for all $F\in C_{t}(H^{2n}(K)).$ This means that
$\Delta_{K}^{n}$ on the domain $\mathcal{H}^{2n}(K_{\mathbb{C}},\nu_{t})$ is
an extension of a self-adjoint operator, and one cannot have a nontrivial
symmetric extension of a self-adjoint operator.

Now, to prove that $\Delta_{K}^{n}$ is symmetric on $\mathcal{H}%
^{2n}(K_{\mathbb{C}},\nu_{t}),$ we write out the inner product in polar
coordinates. It is convenient in this calculation to have the $K$-factor on
the right, so we write $g=xe^{iY}=e^{iY^{\prime}}x,$ where $Y^{\prime
}=\mathrm{Ad}_{x}(Y).$ Then we have (by (\ref{dg.form}))
\begin{equation}
\left\langle \Delta_{K}^{n}F_{1},F_{2}\right\rangle _{L^{2}(K_{\mathbb{C}}%
,\nu_{t})}=\int_{\mathfrak{k}}\int_{K}[\Delta_{K}^{n}\bar{F}_{1}%
](e^{iY^{\prime}}x)F_{2}(e^{iY^{\prime}}x)\,dx\,\frac{\nu_{t}(e^{iY^{\prime}%
})}{\Phi(Y^{\prime})^{2}}dY^{\prime}, \label{reverse}%
\end{equation}
since $\nu_{t}$ is bi-$K$-invariant. In the expression $[\Delta_{K}^{n}\bar
{F}_{1}](e^{iY^{\prime}}x)$ we are thinking of $\Delta_{K}^{n}$ as a
left-invariant differential operator on $K_{\mathbb{C}},$ applied to the
function $\bar{F}_{1}$ and evaluated at the point $e^{iY^{\prime}}x.$ However,
from (\ref{lap.kc}) we see that this is the same as applying $\Delta_{K}$ in
the $x$-variable with $Y^{\prime}$ fixed. Then, since $F_{1}$ and $F_{2}$ are
smooth and since $\Delta_{K}^{n}$ is symmetric on $C^{\infty}(K),$ we can
integrate by parts in the inner integral to get%
\begin{align*}
\left\langle \Delta_{K}^{n}F_{1},F_{2}\right\rangle _{L^{2}(K_{\mathbb{C}}%
,\nu_{t})}  &  =\int_{\mathfrak{k}}\int_{K}\bar{F}_{1}(e^{iY^{\prime}%
}x)[\Delta_{K}^{n}F_{2}](e^{iY^{\prime}}x)\,dx\,\frac{\nu_{t}(e^{iY^{\prime}%
}x)}{\Phi(Y^{\prime})^{2}}dY^{\prime}\\
&  =\left\langle F_{1},\Delta_{K}^{n}F_{2}\right\rangle _{L^{2}(K_{\mathbb{C}%
},\nu_{t})}.
\end{align*}
\end{proof}

We now wish to compute the reproducing kernel for $\mathcal{H}^{2n}%
(K_{\mathbb{C}},\nu_{t}).$ (See \cite{mexnotes} for generalities on
reproducing kernels.) Let us first recall the situation concerning the
reproducing kernel for $\mathcal{H}L^{2}(K_{\mathbb{C}},\nu_{t}),$ since the
reproducing kernel for $\mathcal{H}^{2n}(K_{\mathbb{C}},\nu_{t})$ is computed
by relating it to the reproducing kernel for $\mathcal{H}L^{2}(K_{\mathbb{C}%
},\nu_{t}).$ For each $g\in K_{\mathbb{C}}$, the pointwise evaluation map
$F\rightarrow F(g)$ is a continuous linear functional on $\mathcal{H}%
L^{2}(K_{\mathbb{C}},\nu_{t}).$ Thus, by the Riesz representation theorem,
there exists a unique vector $\chi_{g}\in\mathcal{H}L^{2}(K_{\mathbb{C}}%
,\nu_{t})$ such that%
\begin{equation}
F(g)=\left\langle \chi_{g},F\right\rangle _{L^{2}(K_{\mathbb{C}},\nu_{t})}
\label{coherent1}%
\end{equation}
for all $F\in\mathcal{H}L^{2}(K_{\mathbb{C}},\nu_{t}).$ (We adopt the
convention that the inner product be linear in the second factor.) The vector
$\chi_{g}$ is called the \textbf{coherent state} for $\mathcal{H}%
L^{2}(K_{\mathbb{C}},\nu_{t})$ at the point $g.$ This state also depends on
$t,$ but we have suppressed this dependence in the notation.

In \cite{H1} it is shown that%
\begin{equation}
\chi_{g}(h)=\rho_{2t}(g^{\ast}h). \label{coherent2}%
\end{equation}
Here $\rho_{t}$ is the heat kernel at the identity for $K,$ analytically
continued from $K$ to $K_{\mathbb{C}}$ [Section 4 of \cite{H1}], and the map
$g\rightarrow g^{\ast}$ is the unique antiholomorphic anti-involution of
$K_{\mathbb{C}}$ such that $x^{\ast}=x^{-1}$ for $x\in K.$ If $K=\mathsf{U}%
(n)$ and $K_{\mathbb{C}}=\mathsf{GL}(n;\mathbb{C})$ then $g^{\ast}$ is simply
the usual matrix adjoint. In polar coordinates we have $(xe^{iY})^{\ast
}=e^{iY}x^{-1}.$ It can be shown that $\overline{\rho_{2t}(g^{\ast}h)}%
=\rho_{2t}(gh^{\ast})$ for all $g\in K_{\mathbb{C}}.$ Thus (\ref{coherent1})
and (\ref{coherent2}) become%
\[
F(g)=\int_{K_{\mathbb{C}}}\rho_{2t}(gh^{\ast})F(h)\nu_{t}(h)\,dh.
\]
The function
\[
k_{t}(g,h):=\overline{\chi_{g}(h)}=\rho_{2t}(gh^{\ast})
\]
is called the \textbf{reproducing kernel} for $\mathcal{H}L^{2}(K_{\mathbb{C}%
},\nu_{t}).$

The norm of the pointwise evaluation functional is equal to the norm of the
corresponding coherent state, which can be computed as%
\[
\left\Vert \chi_{g}\right\Vert ^{2}=\left\langle \chi_{g},\chi_{g}%
\right\rangle _{L^{2}(K_{\mathbb{C}},\nu_{t})}=k_{t}(g,g)=\rho_{2t}(gg^{\ast
}).
\]
The pointwise bounds (\ref{kc.bounds}) from \cite{H3} are obtained by
estimating the behavior of the quantity $\rho_{2t}(gg^{\ast}).$

We now turn to the case of the Sobolev spaces. We consider on $H^{2n}(K)$ the
inner product given by%
\begin{equation}
\left\langle f_{1},f_{2}\right\rangle _{H^{2n}(K)}=\left\langle (cI-\Delta
_{K})^{n}f_{1},(cI-\Delta_{K})^{n}f_{2}\right\rangle _{L^{2}(K)},
\label{hn.inner}%
\end{equation}
where $c$ is a positive constant whose value will be chosen later. (Recall
that our Laplacian is negative.) Different positive values of $c$ give
equivalent inner products on $H^{2n}(K),$ and for any $c,$ the inner product
(\ref{hn.inner}) is equivalent to the inner product $\left\langle f_{1}%
,f_{2}\right\rangle _{L^{2}(K)}+\left\langle \Delta_{K}^{n}f_{1},\Delta
_{K}^{n}f_{2}\right\rangle _{L^{2}(K)}.$ The Sobolev space $H^{2n}(K)$ is
complete in the inner product (\ref{hn.inner}), because $\Delta_{K}^{n}$ is
closed on $H^{2n}(K).$

We consider the image under $C_{t}$ of this space, which is denoted
$\mathcal{H}^{2n}(K_{\mathbb{C}},\nu_{t})$ and is characterized in Theorem
\ref{sobolev.thm1}. The map $C_{t}:H^{2n}(K)\rightarrow\mathcal{H}%
^{2n}(K_{\mathbb{C}},\nu_{t})$ will be isometric if we use on $\mathcal{H}%
^{2n}(K_{\mathbb{C}},\nu_{t})$ the inner product%
\begin{equation}
\left\langle F_{1},F_{2}\right\rangle _{\mathcal{H}^{2n}(K_{\mathbb{C}}%
,\nu_{t})}=\left\langle (cI-\Delta_{K})^{n}F_{1},(cI-\Delta_{K})^{n}%
F_{2}\right\rangle _{L^{2}(K_{\mathbb{C}},\nu_{t})}. \label{hnc.inner}%
\end{equation}
The holomorphic Sobolev space $\mathcal{H}^{2n}(K_{\mathbb{C}},\nu_{t})$ is
then complete with respect to the inner product (\ref{hnc.inner}).

We now define the \textbf{coherent states} for $\mathcal{H}^{2n}%
(K_{\mathbb{C}},\nu_{t})$ to be the elements $\chi_{g}^{2n}$ of $\mathcal{H}%
^{2n}(K_{\mathbb{C}},\nu_{t})$ such that%
\[
F(g)=\left\langle \chi_{g}^{2n},F\right\rangle _{\mathcal{H}^{2n}%
(K_{\mathbb{C}},\nu_{t})}.
\]
The \textbf{reproducing kernel} for $\mathcal{H}^{2n}(K_{\mathbb{C}},\nu_{t})$
is then defined as the function $k_{t}^{2n}(g,h)=\overline{\chi_{g}^{2n}(h)}.$

\begin{proposition}
\label{repro.prop}The reproducing kernel for $\mathcal{H}^{2n}(K_{\mathbb{C}%
},\nu_{t})$ is the function $k_{t}^{2n}(g,h)$ given by%
\[
k_{t}^{2n}(g,h)=[(cI-\Delta_{K})^{-2n}\rho_{2t}](gh^{\ast})
\]
which may be computed as%
\begin{equation}
k_{t}^{2n}(g,h)=\frac{1}{(2n-1)!}\int_{0}^{\infty}s^{2n-1}e^{-cs}\rho
_{2(t+s)}(gh^{\ast})\,ds. \label{kn.form}%
\end{equation}
Thus, for all $g\in K_{\mathbb{C}}$ and all $F\in\mathcal{H}^{2n}%
(K_{\mathbb{C}},\nu_{t})$ we have%
\begin{equation}
|F(g)|^{2}\leq\left\|  F\right\|  _{\mathcal{H}^{2n}(K_{\mathbb{C}},\nu_{t}%
)}^{2}k_{t}^{2n}(g,g). \label{knn}%
\end{equation}
\end{proposition}

Note that since $\Delta_{K}$ is a non-positive self-adjoint operator on
$L^{2}(K)$ and also on $\mathcal{H}^{2n}(K_{\mathbb{C}},\nu_{t}),$
$(cI-\Delta_{K})^{-2n}$ is a bounded self-adjoint operator on both $L^{2}(K)$
and $\mathcal{H}^{2n}(K_{\mathbb{C}},\nu_{t}).$ The expression $(cI-\Delta
_{K})^{-2n}\rho_{2t}$ may be interpreted in one of two equivalent ways. We may
think of $\rho_{2t}$ as a function on $K,$ apply $(cI-\Delta_{K})^{-2n},$ and
then analytically to $K_{\mathbb{C}}.$ Alternatively, we may first
analytically continue $\rho_{2t}$ to $K_{\mathbb{C}}$, think of it as an
element of $\mathcal{H}^{2n}(K_{\mathbb{C}},\nu_{t}),$ and then apply
$(cI-\Delta_{K})^{-2n}.$ Since $\Delta_{K}$ (and so also $(cI-\Delta
_{K})^{-2n}$) commutes with analytic continuation, these two views are equivalent.

\begin{proof}
For $F\in\mathcal{H}^{2n}(K_{\mathbb{C}},\nu_{t})\subset\mathcal{H}%
L^{2}(K_{\mathbb{C}},\nu_{t})$ we have
\begin{align*}
F(g)  &  =\left\langle \chi_{g},F\right\rangle _{L^{2}(K_{\mathbb{C}},\nu
_{t})}\\
&  =\left\langle (cI-\Delta_{K})^{n}(cI-\Delta_{K})^{-2n}\chi_{g}%
,(cI-\Delta_{K})^{n}F\right\rangle _{L^{2}(K_{\mathbb{C}},\nu_{t})}\\
&  =\left\langle (cI-\Delta_{K})^{-2n}\chi_{g},F\right\rangle _{\mathcal{H}%
^{2n}(K_{\mathbb{C}},\nu_{t})},
\end{align*}
because $(cI-\Delta_{K})^{-n}$ is a self-adjoint operator on $\mathcal{H}%
L^{2}(K_{\mathbb{C}},\nu_{t}).$ This means that the coherent state for
$\mathcal{H}^{2n}(K_{\mathbb{C}},\nu_{t})$ is $(cI-\Delta_{K})^{-2n}\chi_{g}.$

To compute $(cI-\Delta_{K})^{-2n}$ we use the elementary calculus identity%
\begin{equation}
\frac{1}{a^{2n}}=\frac{1}{(2n-1)!}\int_{0}^{\infty}s^{2n-1}e^{-as}\,ds,\quad
a>0. \label{integral.id}%
\end{equation}
Applying this formally with $a=(cI-\Delta_{K})$ we have%
\begin{equation}
(cI-\Delta_{K})^{-2n}=\frac{1}{(2n-1)!}\int_{0}^{\infty}s^{2n-1}%
e^{-cs}e^{s\Delta_{K}}\,ds. \label{integral.id2}%
\end{equation}
It is not hard to show that this formal argument is correct. Note that since
$\Delta_{K}\leq0,$ the integral on the right-hand side of (\ref{integral.id2})
is absolutely convergent in the operator norm topology.

Now applying this to the function $\rho_{2t}$ and noting that $e^{s\Delta_{K}%
}\rho_{2t}=\rho_{2t+2s}$ (with our normalization of the heat equation) we
obtain%
\begin{equation}
(cI-\Delta_{K})^{-2n}\rho_{2t}=\frac{1}{(2n-1)!}\int_{0}^{\infty}%
s^{2n-1}e^{-cs}\rho_{2(t+s)}\,ds. \label{integral.id3}%
\end{equation}
Here we may initially think of the integral in (\ref{integral.id3}) as taking
values in the Hilbert space $\mathcal{H}L^{2}(K_{\mathbb{C}},\nu_{t}).$
However, the estimates below will show that the integral is convergent
pointwise for all $h$ in $K_{\mathbb{C}}$. (More precisely, the estimates will
show convergence for points of the form $h=g^{\ast}g=e^{2iY}. $ For general
$h\in K_{\mathbb{C}}$ we use the inequality, deduced from the matrix-entry
expansion of $\rho_{t},$ $\left\vert \rho_{t}(xe^{iY})\right\vert
\leq\left\vert \rho_{t}(e^{iY})\right\vert .$)
\end{proof}

\begin{lemma}
\label{at.lem}For all $t>0$ there exists a constant $\alpha_{t}$ such that for
all $\tau>t$ and all $g\in K_{\mathbb{C}}$ we have%
\[
\rho_{2\tau}(gg^{\ast})\leq\alpha_{t}\tau^{(r-d)/2}e^{\left\vert
\delta\right\vert ^{2}\tau}e^{\left\vert Y\right\vert ^{2}/\tau}\Phi(Y).
\]
Here $d$ is the dimension of $K,$ $r=\dim\mathfrak{t}$ is the rank of $K,$
$\delta$ is half the sum of the positive roots, and $\Phi$ is as given in
(\ref{phi.def}).
\end{lemma}

This result is a sharpening of Theorem 2 of \cite{H3}, obtained by estimating
the behavior of the constants $a_{t}$ in that theorem as $t$ tends to
infinity. Assuming this result for the moment, let us complete the proof of
the holomorphic Sobolev embedding theorem (Theorem \ref{embed.thm}).

\begin{proof}
[Proof of Theorem \ref{embed.thm}]We use Proposition \ref{repro.prop} and
apply Lemma \ref{at.lem} with $\tau=t+s$. Since $d=\dim K\geq r=\dim
\mathfrak{t},$ we have $\tau^{(r-d)/2}\leq t^{(r-d)/2}$ for $\tau\geq t$ and
we obtain%
\[
k_{t}^{2n}(g,g)\leq\frac{\alpha_{t}t^{(r-d)/2}e^{t\left\vert \delta\right\vert
^{2}}}{(2n-1)!}\Phi(Y)\int_{0}^{\infty}s^{2n-1}e^{-(c-\left\vert
\delta\right\vert ^{2})s}e^{\left\vert Y\right\vert ^{2}/(s+t)}\,ds.
\]
We now choose $c$ so that $c>\left\vert \delta\right\vert ^{2}.$ Multiplying
and dividing by $e^{\left\vert Y\right\vert ^{2}/t}$ and doing some algebra
gives
\begin{equation}
k_{t}^{2n}(g,g)\leq\frac{\beta_{t}}{(2n-1)!}\Phi(Y)e^{\left\vert Y\right\vert
^{2}/t}\int_{0}^{\infty}s^{2n-1}e^{-Bs}e^{-s\left\vert Y\right\vert
^{2}/t(s+t)}\,ds, \label{kn.bnd2}%
\end{equation}
where $B=c-\left\vert \delta\right\vert ^{2}$ and $\beta_{t}$ is independent
of $g.$

We now divide the integral (\ref{kn.bnd2}) into the region where $s\leq t$ and
the region where $s>t.$ When $s\leq t,$ $s/t(s+t)\geq s/2t^{2}$ and we get%
\begin{align*}
\int_{0}^{t}s^{2n-1}e^{-Bs}e^{-s\left\vert Y\right\vert ^{2}/t(s+t)}ds  &
\leq\int_{0}^{\infty}s^{2n-1}\exp\left\{  -s\left[  B+\frac{\left\vert
Y\right\vert ^{2}}{2t^{2}}\right]  \right\}  ds\\
&  =\frac{(2n-1)!}{(B+\left\vert Y\right\vert ^{2}/2t^{2})^{2n}}%
\end{align*}
by (\ref{integral.id}). When $s>t,$ $s/t(s+t)>s/t(2s)=1/2t$ and we get%
\begin{align*}
\int_{t}^{\infty}s^{2n-1}e^{-Bs}e^{-s\left\vert Y\right\vert ^{2}/t(s+t)}ds
&  \leq e^{-\left\vert Y\right\vert ^{2}/2t}\int_{0}^{\infty}s^{2n-1}%
e^{-Bs}ds\\
&  =\frac{(2n-1)!}{B^{2n}}e^{-\left\vert Y\right\vert ^{2}/2t}.
\end{align*}
Plugging these estimates into (\ref{kn.bnd2}) gives%
\begin{align*}
k_{t}^{2n}(g,g)  &  \leq\beta_{t}\Phi(Y)e^{\left\vert Y\right\vert ^{2}/t}
\left[  \frac{1}{(B+\left\vert Y\right\vert ^{2}/2t^{2})^{2n}}+\frac
{e^{-\left\vert Y\right\vert ^{2}/2t}}{B^{2n}}\right] \\
&  \leq\gamma_{t}\Phi(Y)e^{\left\vert Y\right\vert ^{2}/t}\frac{1}%
{(1+\left\vert Y\right\vert ^{2})^{2n}}.
\end{align*}
This estimate, together with (\ref{knn}), implies the holomorphic Sobolev
embedding theorem, Theorem \ref{embed.thm}.
\end{proof}

It now remains only to prove Lemma \ref{at.lem}.

\begin{proof}
[Proof of Lemma \ref{at.lem}]The paper \cite{H3} establishes the bound%
\begin{equation}
\rho_{2\tau}(gg^{\ast})\leq a_{\tau}e^{\left\vert \delta\right\vert ^{2}\tau
}(4\pi\tau)^{-d/2}\Phi(Y)e^{\left\vert Y\right\vert ^{2}/\tau},\quad
g=xe^{iY}, \label{h3.bound}%
\end{equation}
where $a_{\tau}$ is a quantity independent of $g.$ To establish Lemma
\ref{at.lem} we must show that the optimal constants $a_{\tau}$ can be bounded
by a constant times $\tau^{r/2}$ as $\tau$ tends to infinity. According to
Proposition 3 of \cite{H3} we have%
\[
a_{\tau}\leq\sum_{\gamma\in\bar{C}\cap\Gamma}P\left(  \frac{\left\vert
\gamma\right\vert }{\sqrt{\tau}}\right)  e^{-\left\vert \gamma\right\vert
^{2}/\tau}
\]
where $\bar{C}$ is the closed fundamental Weyl chamber, $P$ is a polynomial,
and $\Gamma\subset\mathfrak{t}$ is the kernel of the exponential mapping,
which is a lattice in $\mathfrak{t}.$ We rewrite this as%
\[
a_{\tau}\leq\sum_{\eta\in\bar{C}\cap(\Gamma/\sqrt{\tau})}P(\left\vert
\eta\right\vert )e^{-\left\vert \eta\right\vert ^{2}}.
\]

If we let $r=\dim\mathfrak{t}$ then it is straightforward to show, using
dominated convergence, that%
\begin{equation}
\lim_{\tau\rightarrow\infty}\frac{1}{\tau^{r/2}}\sum_{\eta\in\bar{C}%
\cap(\Gamma/\sqrt{\tau})}P(\left|  \eta\right|  )e^{-\left|  \eta\right|
^{2}}=\frac{1}{A}\int_{\bar{C}}P(\left|  x\right|  )e^{-\left|  x\right|
^{2}}dx, \label{int.bound}%
\end{equation}
where $A$ is the volume of a fundamental domain in $\Gamma.$ (Approximate the
integrand on the right-hand side of (\ref{int.bound}) by a function that is
constant on each cell of the lattice $\Gamma.$) Thus, the left-hand side of
(\ref{int.bound}) is bounded as $\tau$ tends to infinity. This means that on
each interval of the form $[t,\infty)$ we will have $a_{\tau}$ bounded by a
constant (depending on $t$) times $\tau^{r/2}.$ This (together with
(\ref{h3.bound})) gives the estimate in Lemma \ref{at.lem}.
\end{proof}

\section{Holomorphic Sobolev spaces and Toeplitz operators\label{toeplitz.sec}%
}

Our goal in this section is to show that the holomorphic Sobolev space
$\mathcal{H}^{2n}(K_{\mathbb{C}},\nu_{t})$ (Definition \ref{sobolev.def}) can
described as a holomorphic $L^{2}$ space in which the measure is the heat
kernel measure $\nu_{t}(g)\,dg$ multiplied by the additional factor
$(1+\left|  Y\right|  ^{2})^{2n}.$ As explained at the end of Section
\ref{intro.sec}, this result and the holomorphic Sobolev embedding theorem
(proved in the previous section) together imply Theorem \ref{main.thm},
characterizing the image under $C_{t}$ of $C^{\infty}(K).$

Our strategy is as follows. By a fairly simple integration-by-parts argument,
we will obtain a positive function $\phi_{2n}$ with the property that for
sufficiently nice holomorphic functions $F_{1}$ and $F_{2}$ we have%
\[
\left\langle (cI-\Delta_{K}^{{}})^{n}F,(cI-\Delta_{K}^{{}})^{n}F\right\rangle
_{L^{2}(K_{\mathbb{C}},\nu_{t})}=\int_{K_{\mathbb{C}}}\left|  F(g)\right|
^{2}\phi_{2n}(g)\nu_{t}(g)\,dg.
\]
This means that (for sufficiently nice functions) the inner product on
$\mathcal{H}^{2n}(K_{\mathbb{C}},\nu_{t})$ coincides with the inner product on
$\mathcal{H}L^{2}(K_{\mathbb{C}},\phi_{2n}\nu_{t}).$ It is then not difficult
to show that the Hilbert space $\mathcal{H}^{2n}(K_{\mathbb{C}},\nu_{t})$
coincides with the Hilbert space $\mathcal{H}L^{2}(K_{\mathbb{C}},\phi_{2n}%
\nu_{t}).$ The proof will then be completed by showing that the function
$\phi_{2n}(g)$ has the same behavior at infinity as the function $(1+\left|
Y\right|  ^{2})^{2n}.$

\begin{definition}
Let $\phi$ be a complex-valued measurable function on $K_{\mathbb{C}}$ (not
necessarily holomorphic). Consider the subspace $\mathcal{D}_{\phi}$ of
$\mathcal{H}L^{2}(K_{\mathbb{C}},\nu_{t})$ given by%
\[
\mathcal{D}_{\phi}=\{F\in\mathcal{H}L^{2}(K_{\mathbb{C}},\nu_{t})|\phi F\in
L^{2}(K_{\mathbb{C}},\nu_{t})\}.
\]
Then define the \textbf{Toeplitz operator} $T_{\phi}$ to be the (possibly
unbounded) operator on $\mathcal{H}L^{2}(K_{\mathbb{C}},\nu_{t})$ with domain
$\mathcal{D}_{\phi}$ given by%
\[
T_{\phi}(F)=P_{t}(\phi F),\quad F\in\mathcal{D}_{\phi}.
\]
Here $P_{t}$ is the orthogonal projection operator from $L^{2}(K_{\mathbb{C}%
},\nu_{t})$ onto the closed subspace $\mathcal{H}L^{2}(K_{\mathbb{C}},\nu
_{t})$. The function $\phi$ is called the \textbf{Toeplitz symbol} of the
Toeplitz operator $T_{\phi}.$
\end{definition}

This means that on $\mathcal{D}_{\phi},$ $T_{\phi}$ is equal to $P_{t}M_{\phi
},$ where $M_{\phi}$ denotes multiplication by $\phi.$ If $\phi$ is bounded
then $\mathcal{D}_{\phi}=\mathcal{H}L^{2}(K_{\mathbb{C}},\nu_{t})$ and
$T_{\phi}$ is a bounded operator. In general, $T_{\phi}$ may not be densely
defined in $\mathcal{H}L^{2}(K_{\mathbb{C}},\nu_{t}),$ though it will be
densely defined for the examples we will consider. It is possible that two
different symbols could give rise to the same Toeplitz operator.

\begin{proposition}
\label{toep.easy}For any $F_{1}\in\mathcal{H}L^{2}(K_{\mathbb{C}},\nu_{t})$
and $F_{2}\in\mathcal{D}_{\phi}$ we have%
\[
\left\langle F_{1},T_{\phi}F_{2}\right\rangle _{L^{2}(K_{\mathbb{C}},\nu_{t}%
)}=\int_{K_{\mathbb{C}}}\bar{F}_{1}(g)\phi(g)F_{2}(g)\nu_{t}(g)\,dg.
\]
\end{proposition}

\begin{proof}
Since $P_{t}$ is self-adjoint on $L^{2}(K_{\mathbb{C}},\nu_{t})$ and since
$P_{t}F_{1}=F_{1},$ we have $\left\langle F_{1},T_{\phi}F_{2}\right\rangle
=\left\langle F_{1},P_{t}M_{\phi}F_{2}\right\rangle =\left\langle
F_{1},M_{\phi}F_{2}\right\rangle .$
\end{proof}

Our goal is to express each left-invariant differential operator $A$ acting on
$\mathcal{H}L^{2}(K_{\mathbb{C}},\nu_{t})$ as a Toeplitz operator with some
symbol $\phi_{A}.$ The function $\phi_{2n}$ in the second paragraph of this
section will then be the Toeplitz symbol of the operator $(cI-\Delta_{K}%
)^{2n}.$

We consider the universal enveloping algebra $U(\mathfrak{k})$ of
$\mathfrak{k}$ (with complex coefficients). Then $U(\mathfrak{k})$ is
isomorphic to the algebra of left-invariant differential operators on $K$
(with complex coefficients). Each element of $U(\mathfrak{k})$ can also be
regarded as a left-invariant differential operator on $K_{\mathbb{C}}$ (as in
the case of $\Delta_{K}$).

We then consider the universal enveloping algebra $U(\mathfrak{k}_{\mathbb{C}%
})$ of $\mathfrak{k}_{\mathbb{C}}.$ Here we regard $\mathfrak{k}_{\mathbb{C}}
$ as a \textit{real} Lie algebra, but we use complex coefficients in
constructing $U(\mathfrak{k}_{\mathbb{C}}).$ Thus $U(\mathfrak{k}_{\mathbb{C}%
})$ is isomorphic to the algebra of left-invariant differential operators on
$K_{\mathbb{C}}$ (with complex coefficients). So we now introduce the notation
$J:\mathfrak{k}_{\mathbb{C}}\rightarrow\mathfrak{k}_{\mathbb{C}}$ for the
``multiplication by $i$'' map on $\mathfrak{k}_{\mathbb{C}}.$ So for
$X\in\mathfrak{k}$, we have two different objects, $JX$ and $iX.$ Viewed as
differential operators, these satisfy%
\begin{align*}
JX\phi(g)  &  =\left.  \frac{d}{dt}\phi(ge^{iX})\right|  _{t=0}\\
iX\phi(g)  &  =\left.  i\frac{d}{dt}\phi(ge^{X})\right|  _{t=0},
\end{align*}
for $\phi\in C^{\infty}(K_{\mathbb{C}}).$ If $\phi$ happens to be holomorphic,
$JX$ and $iX$ will coincide. (In the same way, the operators $\partial
/\partial y$ and $i\partial/\partial x$ on $\mathbb{C}$ are not equal, but
they do agree on holomorphic functions.)

\begin{proposition}
There exists a unique homomorphism $\Psi:U(\mathfrak{k})\rightarrow
U(\mathfrak{k}_{\mathbb{C}})$ such that $\Psi(1)=1$ and such that%
\[
\Psi(X)=\frac{1}{2}(X+iJX)
\]
for all $X\in\mathfrak{k}.$
\end{proposition}

\begin{proof}
In light of standard properties of universal enveloping algebras, it suffices
to compute that%
\begin{align*}
\left[  \frac{1}{2}(X+iJX),\frac{1}{2}(Y+iJY)\right]   &  =\frac{1}%
{4}([X,Y]+i[JX,Y]+i[X,JY]-[JX,JY])\\
&  =\frac{1}{4}([X,Y]+iJ[X,Y]+iJ[X,Y]-J^{2}[X,Y])\\
&  =\frac{1}{2}([X,Y]+iJ[X,Y]).
\end{align*}
That is to say, the map $X\rightarrow\frac{1}{2}(X+iJX)$ is a Lie algebra
homomorphism. Here we use that $J^{2}=-I$ and that the bracket on
$\mathfrak{k}_{\mathbb{C}}$ is $J$-linear.
\end{proof}

Suppose that $F$ is holomorphic, so that $\bar{F}$ is antiholomorphic. Then
$JX\bar{F}=-iX\bar{F}.$ From this it follows that%
\[
\frac{1}{2}(X+iJX)\bar{F}=X\bar{F}
\]
for all $X\in\mathfrak{k}$ and therefore%
\begin{equation}
A\bar{F}=\Psi(A)\bar{F},\quad F\in\mathcal{H}(K_{\mathbb{C}})
\label{apsi.holo}%
\end{equation}
for all $A\in U(\mathfrak{k}).$ We will make use of this below.

\begin{definition}
Let $\mathcal{F}\subset\mathcal{H}(K_{\mathbb{C}})$ denote the space of finite
linear combinations of holomorphic matrix entries, that is, the space of
finite linear combinations of functions of the form $F(g)=\mathrm{trace}%
(\pi(g)B),$ where $\pi$ is a finite-dimensional irreducible holomorphic
representation of $K_{\mathbb{C}}$ acting on some vector space $V$ and where
$B$ is a linear operator on $V.$
\end{definition}

For each representation $\pi,$ the space of matrix entries is
finite-dimensional and invariant under all left-invariant differential
operators. In particular, if $F$ is a matrix entry for $\pi,$ then $\Delta
_{K}F=-\lambda_{\pi}F,$ where $\lambda_{\pi}$ is a non-negative constant
depending on $\pi$ but not on $B.$ It is shown in \cite{H1} that
$\mathcal{H}L^{2}(K_{\mathbb{C}},\nu_{t})$ is the orthogonal direct sum of the
spaces of matrix entries, as $\pi$ ranges over the equivalence classes of
irreducible representations of $K_{\mathbb{C}}.$ From these observations it
follows that $\mathcal{F}$ is a core for $(cI-\Delta_{K})^{n},$ for each $n. $

\begin{theorem}
\label{toep.thm}Fix $A\in U(\mathfrak{k}).$ Let $\phi_{A}$ be the function on
$K_{\mathbb{C}}$ given by%
\[
\phi_{A}=\frac{\Psi(A)\nu_{t}}{\nu_{t}},
\]
Then $\mathcal{F}\subset\mathcal{D}_{\phi_{A}}$ and for all $F\in\mathcal{F}$
we have%
\begin{equation}
AF=T_{\phi_{A}}F, \label{af.phif}%
\end{equation}
where on the left-hand side of (\ref{af.phif}), $A$ is regarded as a
left-invariant differential operator on $K_{\mathbb{C}}.$
\end{theorem}

Note that we are \textit{not} asserting that $A=T_{\phi_{A}}$ (with equality
of domains), but only $A=T_{\phi_{A}}$ on the subspace $\mathcal{F} $ of the
domain of $T_{\phi}.$ Equality of domains probably does not hold in general,
although we will see eventually that it is true if $A=(cI-\Delta_{K})^{n}$
(Remark \ref{domain.eq}). Some cases of this result were announced in
\cite{qmphase}.

\begin{proof}
Let us assume for the moment that $\mathcal{F}\subset\mathcal{D}_{\phi_{A}}.$
Once this is established, it suffices (by Proposition \ref{toep.easy}) to show
that
\[
\left\langle F_{1},AF_{2}\right\rangle _{L^{2}(K_{\mathbb{C}},\nu_{t}%
)}=\left\langle F_{1},M_{\phi_A}F_{2}\right\rangle _{L^{2}(K_{\mathbb{C}}%
,\nu_{t})}
\]
for all $F_{1},F_{2}\in\mathcal{F}.$ It suffices to consider $A$ of the form
$A=X_{1}\cdots X_{n}$ with $X_{k}\in\mathfrak{k},$ since every element of
$U(\mathfrak{k})$ is a linear combination of elements of this form. We use
integration by parts on $K_{\mathbb{C}}$ in the form%
\begin{equation}
\int_{K_{\mathbb{C}}}\phi(g)(Z\psi)(g)\,dg=-\int_{K_{\mathbb{C}}}%
(Z\phi)(g)\psi(g)\,dg \label{parts}%
\end{equation}
for any $Z\in\mathfrak{k}_{\mathbb{C}}$. This holds for all sufficiently
regular functions $\phi$ and $\psi$ on $K_{\mathbb{C}}$ (not necessarily
holomorphic). (More on the conditions on $\phi$ and $\psi$ below.) Since we
have written this without any complex conjugates, we can extend this by
linearity to complex linear combinations of elements of $\mathfrak{k}%
_{\mathbb{C}}.$ In particular, for any $X\in\mathfrak{k}$ we have%
\begin{equation}
\int_{K_{\mathbb{C}}}\phi(g)((X+iJX)\psi)(g)\,dg=-\int_{K_{\mathbb{C}}%
}((X+iJX)\phi)(g)\psi(g)\,dg. \label{parts2}%
\end{equation}
Note that on the right-hand side we have (still) $X+iJX$ and \textit{not}
$X-iJX.$

Let us now proceed assuming that all necessary integrations by parts are
valid, addressing this issue at the end. For $X_{1},\ldots,X_{n}%
\in\mathfrak{k}$ we have%
\begin{equation}
\left\langle F_{1},X_{1}\cdots X_{n}F_{2}\right\rangle _{L^{2}(K_{\mathbb{C}%
},\nu_{t})}=\int_{K_{\mathbb{C}}}\bar{F}_{1}(g)(X_{1}\cdots X_{n}F_{2}%
)(g)\nu_{t}(g)\,dg. \label{step1}%
\end{equation}
Since $\nu_{t}(g)$ is bi-$K$-invariant, it is annihilated by each $X_{k}.$
Thus when we integrate by parts, the terms with $X_{k}$ hitting on $\nu_{t}$
are zero and we get%
\begin{align}
&  \left\langle F_{1},X_{1}\cdots X_{n}F_{2}\right\rangle _{L^{2}%
(K_{\mathbb{C}},\nu_{t})}\nonumber\\
&  =(-1)^{n}\int_{K_{\mathbb{C}}}(X_{n}\cdots X_{1}\bar{F}_{1}(g))F_{2}%
(g)\nu_{t}(g)\,dg\label{step2}\\
&  =(-1)^{n}\frac{1}{2^{n}}\int_{K_{\mathbb{C}}}[(X_{n}+iJX_{n})\cdots
(X_{1}+iJX_{1})\bar{F}_{1}](g)F_{2}(g)\nu_{t}(g)\,dg, \label{step3}%
\end{align}
by (\ref{apsi.holo}). We now integrate by parts a second time. When we do so,
the terms where $X_{k}+iJX_{k}$ hit $F_{2}$ are zero, since $F_{2}$ is
holomorphic. Thus we get%
\begin{equation}
\left\langle F_{1},X_{1}\cdots X_{k}F_{2}\right\rangle _{L^{2}(K_{\mathbb{C}%
},\nu_{t})}=\frac{1}{2^{n}}\int_{K_{\mathbb{C}}}\bar{F}_{1}(g)F_{2}%
(g)[(X_{1}+iJX_{1})\cdots(X_{n}+iJX_{n})\nu_{t}](g)\,dg. \label{step4}%
\end{equation}
Multiplying and dividing by $\nu_{t}(g)$ we get%
\[
\left\langle F_{1},X_{1}\cdots X_{n}F_2\right\rangle _{L^{2}(K_{\mathbb{C}}%
,\nu_{t})}=\left\langle F_{1},\frac{\Psi(A)\nu_{t}}{\nu_{t}}F_{2}\right\rangle
_{L^{2}(K_{\mathbb{C}},\nu_{t})}=\left\langle F_{1},\phi_{A}F_{2}\right\rangle
_{L^{2}(K_{\mathbb{C}},\nu_{t})},
\]
which is what we wanted to show.

The heart of the proof of Theorem \ref{toep.thm} is the integration by parts
in the previous paragraph. It remains only to address two technical issues:
showing that $\mathcal{F}\subset\mathcal{D}_{\phi_{A}}$ and showing that the
boundary terms in the integration by parts vanish. We sketch the arguments
here and provide more details in Section \ref{int.sec}. By writing out what
the left-invariant differential operator $\Psi(A)$ looks like in polar
coordinates and by using the explicit formula for $\nu_{t},$ it is not hard to
show that $\Psi(A)\nu_{t}$ behaves at worst like $e^{-\left\vert Y\right\vert
^{2}/t}$ times a function with exponential growth in $Y.$ Thus $\phi_{A}%
=\Psi(A)\nu_{t}/\nu_{t}$ will have at most exponential growth in $Y.$ Since
the holomorphic matrix entries also have at most exponential growth in $Y,$
this (together with the formula (\ref{nut.measure})) shows that $\int
_{K_{\mathbb{C}}}|F(g)\phi_{A}(g)|^{2}\nu_{t}(g)\,dg<\infty$ for any matrix
entry $F.$ That is to say, $\mathcal{F}\subset\mathcal{D}_{\phi_{A}}.$

We must also justify two integrations by parts, one in passing from
(\ref{step1}) to (\ref{step2}) and one in passing from (\ref{step3}) to
(\ref{step4}). The first of these involves only differentiation in the
$K$-directions. If we write out the integral in reverse polar coordinates as
in (\ref{reverse}), the integration by parts will be only in the $K$-integral,
where there are no boundary terms to worry about. So we need only worry about
the passage from (\ref{step3}) to (\ref{step4}). We use the following
criterion for applying (\ref{parts2}): in polar coordinates, $\psi,$ along
with its partial derivatives in the $Y$ variable, should have at most
exponential growth in $Y$ (with estimates uniform in $x$), while $\phi,$ along
with its partial derivatives in the $Y$ variable, should have faster than
exponential decay. (See Section \ref{int.sec} for a justification of this
condition.) As we do successive integrations by parts to pass from
(\ref{step3}) to (\ref{step4}) we will apply this criterion with%
\begin{equation}
\psi(g)=[(X_{k-1}+iJX_{k-1})\cdots(X_{1}+iJX_{1})\bar{F}_{1}](g)
\label{psi.form}%
\end{equation}
and%
\begin{equation}
\phi(g)=F_{2}(g)[(X_{k+1}+iJX_{k+1})\cdots(X_{n}+iJX_{n})\nu_{t}](g).
\label{phi.form}%
\end{equation}
Calculation in polar coordinates will show (Section \ref{int.sec}) that these
functions indeed satisfy the above criterion.
\end{proof}

\begin{proposition}
\label{phin.poly}For any positive integer $n,$ let $\phi_{n}$ denote the
Toeplitz symbol of the operator $(c-\Delta_{K})^{n},$ namely,%
\begin{equation}
\phi_{n}=\frac{\Psi((c-\Delta_{K})^{n})\nu_{t}}{\nu_{t}}. \label{phin.def}%
\end{equation}
Then
\[
\phi_{n}(xe^{iY})=p_{n,c,t}(\left\vert Y\right\vert ^{2}),
\]
where $p_{n,c,t}$ is a polynomial of degree $n.$ Furthermore, for each $n$ and
$t,$ $\phi_{n}$ is a positive function for all sufficiently large values of
$c.$
\end{proposition}

\begin{proof}
Since $X_{k}$ commutes with $JX_{k},$ we have%
\[
\Psi(X_{k}^{2})=\frac{1}{4}(X_{k}+iJX_{k})^{2}=\frac{1}{4}(X_{k}^{2}%
+2i(JX_{k})X_{k}-(JX_{k})^{2}).
\]
Now, $\nu_{t}$ satisfies the differential equation $d\nu_{t}/dt=\frac{1}%
{4}\sum_{k=1}^{d}(JX_{k})^{2}\nu_{t}.$ Furthermore, $\nu_{t}$ is
bi-$K$-invariant and therefore annihilated by each $X_{k}.$ Thus%
\[
\Psi\left(  \Delta_{K}\right)  \nu_{t}=-\frac{1}{4}\sum_{k=1}^{d}(JX_{k}%
)^{2}\nu_{t}=-\frac{d\nu_{t}}{dt}.
\]
Since $t$ derivatives commute with spatial derivatives we then have%
\begin{equation}
\Psi((c-\Delta_{K})^{n})\nu_{t}=\left(  c+\frac{d}{dt}\right)  ^{n}\nu
_{t}=\sum_{k=0}^{n}\binom{n}{k}c^{n-k}\left(  \frac{d}{dt}\right)  ^{k}\nu
_{t}. \label{c.sum}%
\end{equation}

Looking at the formula for $\nu_{t}$ in (\ref{nut.function}) and
(\ref{ct.def}), we see that repeated applications of the operator $d/dt$ to
$\nu_{t}$ will give back $\nu_{t}$ itself multiplied by a sum of terms of the
form $t^{-a}\left\vert Y\right\vert ^{2b}$, with coefficients involving
$\left\vert \delta\right\vert ^{2}$ and $d.$ (Here $a$ and $b$ are
non-negative integers and the result may be proved by induction on the number
of derivatives.) The highest power of $\left\vert Y\right\vert ^{2}$ that will
arise in computing (\ref{c.sum}) is $\left\vert Y\right\vert ^{2n}. $ This
establishes that $\phi_{n}=[(c+d/dt)^{n}\nu_{t}]/\nu_{t}$ is a polynomial in
$\left\vert Y\right\vert ^{2}$ of degree $n.$

To establish the positivity of $\phi_{n}$ , let us think about the coefficient
of $\left\vert Y\right\vert ^{2l}$ in the computation of $\phi_{n}.$ From the
$k=l$ term in (\ref{c.sum}) we get%
\begin{equation}
c^{n-l}\left[  \frac{\left\vert Y\right\vert ^{2l}}{t^{2l}}+\text{lower powers
of }\left\vert Y\right\vert ^{2}\right]  . \label{kl.term}%
\end{equation}
From all terms in (\ref{c.sum}) with $k<l$, we get powers of $\left\vert
Y\right\vert ^{2}$ lower than $\left\vert Y\right\vert ^{2l}.$ From terms in
(\ref{c.sum}) with $k>l,$ we may get terms involving $\left\vert Y\right\vert
^{2l},$ but they will be multiplied by a lower power of $c$ than in
(\ref{kl.term}). We see, then, that for each fixed value of $n$ and $t,$ the
coefficient of $\left\vert Y\right\vert ^{2l}$ in $\phi_{n}$ will be a
polynomial in $c$ of degree $n-l$ with positive leading term. Thus for all
sufficiently large values of $c,$ every power of $\left\vert Y\right\vert
^{2}$ in the expression for $\phi_{n}$ will have a positive coefficient and
$\phi_{n}$ will therefore be positive.
\end{proof}

\begin{proposition}
\label{phin.l2}Choose $c$ large enough that the function $\phi_{2n}$ in
Proposition \ref{phin.poly} is positive. Then the holomorphic Sobolev space
$\mathcal{H}^{2n}(K_{\mathbb{C}},\nu_{t})$ coincides with the Hilbert space
$\mathcal{H}L^{2}(K_{\mathbb{C}},\phi_{2n}(g)\nu_{t}(g)\,dg).$
\end{proposition}

Since, by Proposition \ref{phin.poly}, $\phi_{2n}(g)$ has the same behavior at
infinity as $(1+\left|  Y\right|  ^{2})^{2n},$ Proposition \ref{phin.l2}
implies Theorem \ref{sobolev.thm2}.

\begin{proof}
For $F_{1},F_{2}\in\mathcal{F}$ we have%
\begin{align*}
\left\langle F_{1},F_{2}\right\rangle _{\mathcal{H}^{2n}(K_{\mathbb{C}}%
,\nu_{t})}  &  =\left\langle (c-\Delta_{K})^{n}F_{1},(c-\Delta_{K})^{n}%
F_{2}\right\rangle _{L^{2}(K_{\mathbb{C}},\nu_{t})}\\
&  =\left\langle F_{1},(c-\Delta_{K})^{2n}F_{2}\right\rangle _{L^{2}%
(K_{\mathbb{C}},\nu_{t})}\\
&  =\left\langle F_{1},T_{\phi_{2n}}F_{2}\right\rangle _{L^{2}(K_{\mathbb{C}%
},\nu_{t})}\\
&  =\left\langle F_{1},\phi_{2n}F_{2}\right\rangle _{L^{2}(K_{\mathbb{C}}%
,\nu_{t})}.
\end{align*}
(We have used Proposition \ref{toep.easy} in the last equality.) The last
expression is nothing but the inner product of $F_{1}$ and $F_{2}$ in
$L^{2}(K_{\mathbb{C}},\phi_{2n}\nu_{t}).$ Thus the inner product for
$\mathcal{H}^{2n}(K_{\mathbb{C}},\nu_{t})$ and for $\mathcal{H}L^{2}%
(K_{\mathbb{C}},\phi_{2n}\nu_{t})$ coincide on $\mathcal{F}.$ But (as in the
proof of Theorem \ref{sobolev.thm1}), $\mathcal{F}$ is dense in $\mathcal{H}%
^{2n}(K_{\mathbb{C}},\nu_{t}).$ Furthermore, by the same argument as in the
proof of Lemma 10 in \cite{H1}, $\mathcal{F}$ is dense in $\mathcal{H}%
L^{2}(K_{\mathbb{C}},\phi_{2n}\nu_{t}).$ It then follows that the two Hilbert
spaces $\mathcal{H}^{2n}(K_{\mathbb{C}},\nu_{t})$ and $\mathcal{H}%
L^{2}(K_{\mathbb{C}},\phi_{2n}\nu_{t})$ must coincide.
\end{proof}

\begin{remark}
\label{domain.eq}Proposition \ref{phin.poly} tells us that $\phi_{2n}(g)$ has
the same behavior at infinity as $\phi_{n}(g)^{2}.$ (That is, the Toeplitz
symbol of $(c-\Delta_{K})^{2n}$ has the same behavior at infinity as the
square of the Toeplitz symbol of $(c-\Delta_{K})^{n}.$) From this we see that
$\mathcal{D}_{\phi_{n}}$ is the same space as $\mathcal{H}L^{2}(K_{\mathbb{C}%
},\phi_{2n}\nu_{t}).$ Thus $\mathcal{D}_{\phi_{n}}=\mathcal{H}L^{2}%
(K_{\mathbb{C}},\phi_{2n}\nu_{t})=\mathcal{H}^{2n}(K_{\mathbb{C}},\nu_{t}).$
Thus $(c-\Delta_{K})^{n}=T_{\phi_{n}}$, with equality of domains. For more
general left-invariant operators $A,$ there is no obvious reason that the
symbol of $A^{2}$ should have the same behavior at infinity as the square of
the symbol of $A.$ Thus in general the domain of $A$ may not coincide with
$\mathcal{D}_{\phi_{A}}.$
\end{remark}

\section{Integration by parts and growth of logarithmic
derivatives\label{int.sec}}

In this section we give more details concerning the technical issues in the
proof of Theorem \ref{toep.thm}, namely, justifying integration by parts and
bounding functions of the form $\Psi(A)\nu_{t}/\nu_{t}.$ Our strategy is to
write out left-invariant differential operators on $K_{\mathbb{C}}$ in polar
coordinates. This means that we think of $K_{\mathbb{C}}$ as $K\times
\mathfrak{k}$ by means of polar coordinates and we express everything in terms
of left-invariant vector fields on $K$ and constant coefficient differential
operators on $\mathfrak{k}.$ So we introduce vector fields $\tilde{X}_{k}$ and
$\partial/\partial y_{k}$ on $K_{\mathbb{C}},$ where $\tilde{X}_{k}$ is given
by%
\[
(\tilde{X}_{k}\phi)(xe^{iY})=\left.  \frac{d}{dt}\phi(xe^{tX_{k}}%
e^{iY})\right\vert _{t=0}
\]
and where $\partial/\partial y_{k}$ means partial differentiation in the $Y$
variable with $x$ fixed. Note that both $\tilde{X}_{k}$ and $\partial/\partial
y_{k}$ are left-$K$-invariant operators, but neither is left-$K_{\mathbb{C}}%
$-invariant. Note that since they act in separate variables, the $\tilde
{X}_{k}$'s commutes with the $\partial/\partial y_{l}$'s.

Since the vector fields $\tilde{X}_{k}$ and $\partial/\partial y_{k}$ span the
tangent space at each point, the left-$K_{\mathbb{C}}$-invariant vector fields
$X_{k}$ and $JX_{k}$ can be expressed as linear combinations of these vector
fields with coefficients that are smooth functions on $K_{\mathbb{C}}. $ Since
all the vector fields involved are left-$K$-invariant, the coefficient
functions (in polar coordinates) will depend only on $Y$ and not on $x.$ So we
will have%
\begin{align}
X_{k}  &  =\sum_{l=1}^{d}\left(  a_{kl}(Y)\tilde{X}_{l}+b_{kl}(Y)\frac
{\partial}{\partial y_{l}}\right) \label{xk.form}\\
JX_{k}  &  =\sum_{l=1}^{d}\left(  c_{kl}(Y)\tilde{X}_{l}+d_{kl}(Y)\frac
{\partial}{\partial y_{l}}\right)  . \label{jxk.form}%
\end{align}

The coefficient functions $a_{kl},$ etc., can be computed explicitly by
differentiating the polar coordinates map $(x,Y)\rightarrow xe^{iY}.$ This
calculation is done in \cite{H3} with the result that%
\begin{equation}
\left(
\begin{array}
[c]{cc}%
a(Y) & c(Y)\\
b(Y) & d(Y)
\end{array}
\right)  =\left(  \frac{\sin\mathrm{ad}Y}{\mathrm{ad}Y}\right)  ^{-1}\left(
\begin{array}
[c]{cc}%
\frac{\sin\mathrm{ad}Y}{\mathrm{ad}Y} & \frac{\cos\mathrm{ad}Y-1}%
{\mathrm{ad}Y}\\
\sin\mathrm{ad}Y & \cos\mathrm{ad}Y
\end{array}
\right)  . \label{diff.matrix}%
\end{equation}
Here $\sin\mathrm{ad}Y/\mathrm{ad}Y$ is to be computed using the power series
for the function $\sin z/z,$ which has infinite radius of convergence, and
similarly for $(\cos\mathrm{ad}Y-1)/\mathrm{ad}Y.$ Note that the eigenvalues
of $\mathrm{ad}Y$ are pure imaginary, so the eigenvalues of $\sin
\mathrm{ad}Y/\mathrm{ad}Y$ are of the form $\sin(ia)/(ia)=\sinh a/a,$
$a\in\mathbb{R}.$ This means that $\sin\mathrm{ad}Y/\mathrm{ad}Y$ is
invertible. In (\ref{diff.matrix}), each $d\times d$ block of the
$(2d)\times(2d)$ matrix on the right-hand side is to be multiplied by the
$d\times d$ matrix $(\sin\mathrm{ad}Y/\mathrm{ad}Y)^{-1}.$ Note that the
functions $a,$ $b,$ $c,$ and $d$ have at most linear growth as a function of
$Y.$

\textit{Integration by parts}. We start by verifying the criterion for
integration by parts described in the previous section: $\psi$ and its
$Y$-derivatives should have at most exponential growth; $\phi$ and its
$Y$-derivatives should have faster-than-exponential decay. In justifying the
integration by parts, the $X_{k}$ term is no problem, since then we write out
things in reverse polar coordinates (as in (\ref{reverse})) and the
integration by parts will be purely in the $K$-variable, where there is no
boundary to worry about.

For the $JX_{k}$ term, we write out the integration in polar coordinates,
using (\ref{jxk.form}). This gives%
\begin{align}
\int_{K_{\mathbb{C}}}(JX_{k}\phi)(g)\psi(g)\,dg  &  =\sum_{k=1}^{d}%
\int_{\mathfrak{k}}\int_{K}c_{kl}(Y)(\tilde{X}_{l}\phi)(xe^{iY})\psi
(xe^{iY})\,dx\,\Phi(Y)^{-2}\,dY\nonumber\\
&  +\sum_{k=1}^{d}\int_{K}\int_{\mathfrak{k}}d_{kl}(Y)\frac{\partial\phi
}{\partial y_{l}}(xe^{iY})\psi(xe^{iY})\Phi(Y)^{-2}\,dY\,dx. \label{jxk.int}%
\end{align}
Under our assumptions on $\phi$ and $\psi,$ the integrals are all convergent.
In the first term, we use that $\tilde{X}_{l}$ is skew-symmetric on
$C^{\infty}(K)$ to move $\tilde{X}_{l}$ from $\phi$ onto $\psi$ (with a minus
sign in front). In the second term, we compute the inner integral by
integrating over a cube in $\mathfrak{k}$ and then letting the size of the
cube tend to infinity. In the second term, we apply ordinary Euclidean
integration by parts. This will give three integral terms (from the functions
$d_{kl},$ $\psi,$ and $\Phi^{-2}$ in the integrand) plus a boundary term. Two
of the integral terms are \textquotedblleft divergence\textquotedblright%
\ terms, namely,%
\[
\int_{K}\int_{\mathrm{cube}}\left[  \sum_{k=1}^{d}\left(  \frac{\partial
d_{kl}(Y)}{\partial y_{l}}\Phi^{-2}(Y)+d_{kl}(Y)\frac{\partial\Phi^{-2}%
(Y)}{\partial y_{l}}\right)  \right]  \phi(xe^{iY})\psi(xe^{iY})\,dY\,dx.
\]
Now, the quantity in square brackets must be identically zero, or else
$JX_{k}$ would not be skew-symmetric on $C_{c}^{\infty}(K_{\mathbb{C}}).$ (The
skew-symmetry of $JX_{k}$ is a consequence of the invariance of Haar measure
under right translations on the unimodular group $K_{\mathbb{C}}.$)

We are left, then, with the term we want, namely,%
\[
-\sum_{k=1}^{d}\int_{K}\int_{\mathrm{cube}}\phi(xe^{iY})d_{kl}(Y)\frac
{\partial\psi}{\partial y_{l}}(xe^{iY})\Phi(Y)^{-2}\,dY\,dx
\]
together with a boundary term, namely,%
\[
\sum_{k=1}^{d}\int_{K}\int_{\mathrm{boundary}}d_{kl}(Y)\phi(xe^{iY}%
)\psi(xe^{iY})\Phi(Y)^{-2}\,dY\,dx,
\]
where \textquotedblleft boundary\textquotedblright\ refers to integration over
two opposite faces of the cube (with opposite signs). Since we assume that
$\psi$ and its $Y$-derivatives have at most exponential growth and that $\phi$
has faster-than-exponential decay, we can now let the size of the cube tend to
infinity. The boundary term will drop out in the limit and the remaining term
becomes%
\begin{equation}
-\sum_{k=1}^{d}\int_{K}\int_{\mathfrak{k}}\phi(xe^{iY})d_{kl}(Y)\frac
{\partial\psi}{\partial y_{l}}(xe^{iY})\Phi(Y)^{-2}\,dY\,dx.
\label{second.term}%
\end{equation}

Recall that (\ref{second.term}) is the second term from the right-hand side of
(\ref{jxk.int}). After we integrate by parts in the first term (in the
$x$-variable only) we get%
\[
\int_{K_{\mathbb{C}}}(JX_{k}\phi)(g)\psi(g)\,dg=-\int_{K_{\mathbb{C}}}%
\phi(g)(JX_{k}\psi)(g)\,dg
\]
and our criterion for integration by parts is justified.

It remains to check that the functions $\phi$ and $\psi$ to which we want to
apply integration by parts satisfy the just-obtained criterion. According to
(\ref{psi.form}), $\psi$ is obtained by applying left-invariant derivatives to
the complex conjugate of a matrix entry. Since left-invariant derivatives of
matrix entries are again matrix entries, $\psi$ will have at most exponential
growth. Meanwhile, according to (\ref{phi.form}), $\phi$ is the heat kernel
$\nu_{t},$ with several left-invariant derivatives applied and multiplied by a
matrix entry. As we will show below, left-invariant derivatives of $\nu_{t}$
give back $\nu_{t}$ itself, multiplied by a function with at most exponential
growth. Thus $\phi$ has faster-than-exponential decay.

\textit{Logarithmic-type derivatives}. The entries of (\ref{diff.matrix}) grow
only linearly in $Y.$ However, due to the noncommutative nature of
differentiating matrix-valued functions, it is not altogether evident how the
derivatives of these functions behave. Nevertheless, it is not hard to show
that all the derivatives have at most exponential growth. For example, to
differentiate the expression for $c(Y),$ namely,%
\[
c(Y)=\left(  \frac{\sin\mathrm{ad}Y}{\mathrm{ad}Y}\right)  ^{-1}\frac
{\cos\mathrm{ad}Y-1}{\mathrm{ad}Y}
\]
we use the rules for differentiating products and inverses of matrix-valued
functions, giving, for any smooth path $Y(t),$%
\begin{align*}
\frac{d}{dt}c(Y(t))  &  =-\left(  \frac{\sin\mathrm{ad}Y(t)}{\mathrm{ad}%
Y(t)}\right)  ^{-1}\left[  \frac{d}{dt}\frac{\sin\mathrm{ad}Y(t)}%
{\mathrm{ad}Y(t)}\right]  \left(  \frac{\sin\mathrm{ad}Y(t)}{\mathrm{ad}%
Y(t)}\right)  ^{-1}\\
&  +\left(  \frac{\sin\mathrm{ad}Y(t)}{\mathrm{ad}Y(t)}\right)  ^{-1}\left[
\frac{d}{dt}\frac{\cos\mathrm{ad}Y(t)-1}{\mathrm{ad}Y(t)}\right]  .
\end{align*}
Term-by-term differentiation gives exponential estimates on the two terms in
square brackets.

Since the derivatives of the coefficient functions $a,$ $b,$ $c,$ and $d$ grow
at most exponentially, any left-invariant differential operator on
$K_{\mathbb{C}}$ (built up from products of left-invariant vector fields) will
be expressible in terms of the operators $\tilde{X}_{k}$ and $\partial
/\partial Y_{k}$ with coefficients that have at most exponential growth. When
we apply such an operator to $\nu_{t},$ any term with $\tilde{X}_{k}$ in it is
zero and we get only $Y$-derivatives of $\nu_{t}$, multiplied by at most
exponential coefficient functions. It is easily seen that the $Y$ derivatives
of $\nu_{t}$ can be expressed as $\nu_{t}$ itself, multiplied by functions of
at most exponential growth. Thus functions of the form $\Psi(A)\nu_{t}/\nu
_{t}$ will have at most exponential growth.

\section{Concluding remarks\label{conclude.sec}}

\subsection{Pointwise bounds}

The pointwise bounds (\ref{kc.bounds}) for $\mathcal{H}L^{2}(K_{\mathbb{C}%
},\nu_{t})$---or, equivalently, bounds for the on-diagonal reproducing kernel
$k_{t}(g,g)$---play an essential role in the proof of Theorem \ref{main.thm}.
After all, the holomorphic Sobolev embedding theorem (which provides one
direction of Theorem \ref{main.thm}) is proved by relating the reproducing
kernel $k_{t}^{2n}(g,g)$ for $\mathcal{H}^{2n}(K_{\mathbb{C}},\nu_{t})$ to the
reproducing kernel $k_{t}(g,g)$ for $\mathcal{H}L^{2}(K_{\mathbb{C}},\nu_{t})$
and then using (a slight refinement of) estimates from \cite{H3} for
$k_{t}(g,g).$ (See the expression (\ref{kn.form}) in Proposition
\ref{repro.prop} and Lemma \ref{at.lem}.)

In \cite{H3}, the bounds on $k_{t}(g,g)$ are obtained by using the expression
$k_{t}(g,g)=\rho_{2t}(gg^{\ast})$ from \cite{H1}. This expression, in turn,
depends on the special form of the measure $\nu_{t}(g)\,dg,$ namely, that it
is the heat kernel for $K_{\mathbb{C}}/K.$ In light of the importance of the
bounds (\ref{kc.bounds}), it is worthwhile to compare them to bounds
obtainable by more general means. The elementary argument of \cite{DG} gives
the bound (for $F\in\mathcal{H}L^{2}(K_{\mathbb{C}},\nu(g)~dg)$)%
\begin{equation}
\left\vert F(g)\right\vert ^{2}\leq\left\Vert F\right\Vert _{L^{2}%
(K_{\mathbb{C}},\nu)}^{2}a_{\varepsilon}\sup_{h\in B_{\varepsilon}(g)}\frac
{1}{\nu(h)}, \label{local.bnd1}%
\end{equation}
where $B_{\varepsilon}(g)$ is the ball of radius $\varepsilon$ around $g,$
computed using a left-invariant Riemannian metric on $K_{\mathbb{C}}.$ Here,
$\nu(g)$ is an arbitrary positive, continuous density on $K_{\mathbb{C}}.$ If
we specialize to the case $\nu=\nu_{t},$ it is possible that by using a
\textquotedblleft local gauge transformation\textquotedblright\ as in
\cite{CL}, one could allow $\varepsilon$ to go to zero in the above estimate,
and thereby obtain an estimate of the form%
\begin{equation}
\left\vert F(g)\right\vert ^{2}\leq\left\Vert F\right\Vert _{L^{2}%
(K_{\mathbb{C}},\nu_{t})}^{2}a\frac{1}{\nu_{t}(g)}=A_{t}\frac{e^{\left\vert
Y\right\vert ^{2}/t}}{\Phi(Y)}. \label{local.bnd}%
\end{equation}
This estimate would be \textquotedblleft off\textquotedblright\ from the
actual optimal estimate by a factor of $\Phi(Y)^{2}$ (since (\ref{kc.bounds})
has $\Phi(Y)$ in the numerator rather than the denominator).

Meanwhile, the estimate of Driver \cite{D} together with the averaging lemma
of \cite{H1} imply the bound%
\begin{equation}
\left\vert F(g)\right\vert ^{2}\leq\left\Vert F\right\Vert _{L^{2}%
(K_{\mathbb{C}},\nu_{t})}^{2}A_{t}e^{\left\vert g\right\vert ^{2}/t},
\label{driver.bnd}%
\end{equation}
where $\left\vert g\right\vert ^{2}$ is the distance from the identity to $g$
with respect to a certain left-invariant Riemannian metric on $K_{\mathbb{C}%
}.$ This estimate holds for heat kernel measures on arbitrary complex Lie
groups \cite{DG} and is therefore less dependent on the special structure of
$\nu_{t}$ than the argument in \cite{H3}. On the other hand, it can be shown
that $\left\vert Y\right\vert ^{2}\leq\left\vert g\right\vert ^{2}%
\leq\left\vert Y\right\vert ^{2}+C,$ and therefore (\ref{driver.bnd}) is
equivalent to%
\[
\left\vert F(g)\right\vert ^{2}\leq\left\Vert F\right\Vert _{L^{2}%
(K_{\mathbb{C}},\nu_{t})}^{2}B_{t}e^{\left\vert Y\right\vert ^{2}/t}.
\]
This bound is better by a factor of $\Phi(Y)$ than (\ref{local.bnd}), but
still one factor of $\Phi(Y)$ off from the bounds in (\ref{kc.bounds}).

We see, then, that more general methods leave us exponentially short of the
estimates from \cite{H3} (at least when $K$ is semisimple). To make the proofs
in this paper work, we need bounds that are within a polynomial factor of the
ones in (\ref{kc.bounds}). If there is a way to get bounds similar to
(\ref{kc.bounds}) without using special properties of the density $\nu_{t},$
it would presumably be by working in the right sort of holomorphic local
coordinates about each point $g.$ The coordinate neighborhood about $g$ should
have fixed volume with respect to the \textquotedblleft phase volume
measure\textquotedblright\ $dx\,dY.$ If such coordinates can be constructed,
then one might hope to get estimates involving the reciprocal of the density
of $\nu_{t}(g)\,dg$ with respect to $dx\,dY,$ which is precisely what we have
in (\ref{kc.bounds}). (One would still need some sort of local gauge
transformation to make this work.) By contrast, (\ref{local.bnd1}) and
(\ref{local.bnd}) involve the reciprocal of the density of $\nu_{t}(g)\,dg$
with respect to the Haar measure $dg.$ In the noncommutative case, when the
Haar measure has exponential volume growth, such bounds are not adequate.

\subsection{Connections with the inversion formula}

It is illuminating to think of the results of this paper in comparison to the
inversion formula for the generalized Segal--Bargmann transform given in
\cite{H2}. Consider some $f$ in $L^{2}(K)$ and let $F=C_{t}f.$ Then according
to \cite{H2} we have the inversion formula%
\begin{equation}
f(x)=(2\pi t)^{-d/2}e^{-\left\vert \delta\right\vert ^{2}t/2}\int
_{\mathfrak{k}}F(xe^{iY})e^{-\left\vert Y\right\vert ^{2}/2t}\frac{1}%
{\Phi(Y/2)}\,dY, \label{invert.int}%
\end{equation}
\textit{provided} that the integral is absolutely convergent for all $x.$
(This formula is obtained by writing out the integral in \cite[Eq. (2)]{H2}
explicitly with $p=e^{iY}$ and then making the change of variables $Y^{\prime
}=2Y.$ Compare also Theorem 2.6 in \cite{geoquant}.) For any $f\in L^{2}(K),$
one can recover $f$ from $F$ by restricting the integral in (\ref{invert.int})
to a ball of radius $R$ in $\mathfrak{k}$ and then taking the limit (in the
$L^{2}(K)$ topology) as $R$ tends to infinity.

The question of when the integral in (\ref{invert.int})\ converges is an
important one. The integral cannot be convergent for all $f$ and $x,$ since
$f$ is an arbitrary $L^{2}$ function on $K$ and can equal infinity at some
points. On the other hand, if $f$ is sufficiently smooth (with estimates
depending on the dimension and rank of $K$), Theorem 3 of \cite{H2} shows that
the integral is indeed convergent for all $x\in K.$

The pointwise bounds in \cite{H3} and in this paper reflect this state of
affairs. For general $f$ in $L^{2}(K)$ we have (taking the square root of both
sides of (\ref{kc.bounds}))%
\[
\left\vert F(xe^{iY})\right\vert \leq Be^{\left\vert Y\right\vert ^{2}/2t}%
\Phi(Y)^{1/2}.
\]
Since $\Phi(Y/2)$ has the same asymptotic behavior (up to a constant) as
$\Phi(Y)^{1/2}$, these bounds are polynomially short of what is needed to
guarantee convergence of (\ref{invert.int}). On the other hand, if we assume a
sufficient degree of differentiability for $f,$ then by the holomorphic
Sobolev embedding theorem (Theorem \ref{embed.thm}), we get polynomially
better bounds and thus convergence of (\ref{invert.int}). Therefore, the
holomorphic Sobolev embedding theorem gives an alternative way of proving the
convergence results in Theorem 3 of \cite{H2}. It is interesting to note,
however, that the convergence results of \cite{H2} can be proved by
comparatively soft methods which do not require detailed estimates of the
reproducing kernel. (Compare also the results in \cite{St} extending \cite{H2}
to arbitrary compact symmetric spaces.)

\subsection{Distributions}

In this paper we have considered functions smoother than those in $L^{2}(K),$
either Sobolev spaces or $C^{\infty}(K).$ One could also consider
distributions, which are \textquotedblleft functions\textquotedblright\ less
smooth than those in $L^{2}(K).$ It is easily shown that the transform
$C_{t},$ defined initially on $L^{2}(K),$ can be extended to a map of the
space of distributions on $K$ into the space of holomorphic functions on
$K_{\mathbb{C}}.$ (Compare \cite{fmn1, fmn2}.) We make the following
conjecture concerning the image under $C_{t}$ of the space of distributions.

\begin{conjecture}
Suppose $F$ is a holomorphic function on $K_{\mathbb{C}}.$ Then there exists a
distribution $f$ on $K$ with $F=C_{t}f$ if and only if $F$ satisfies%
\begin{equation}
\left|  F(xe^{iY})\right|  ^{2}\leq A\Phi(Y)e^{|Y|^{2}/t}(1+|Y|^{2})^{2n}
\label{dist.bnd}%
\end{equation}
for some positive integer $n$ and some constant $A.$
\end{conjecture}

This is the same bound as in Theorem \ref{main.thm} except that the factor of
$(1+\left\vert Y\right\vert ^{2})^{2n}$ is in the numerator instead of the
denominator and the bound is required to hold only for \textit{some} $n$
rather than for all $n.$ The most obvious approach to proving such a theorem
would be to consider negative Sobolev spaces $H^{-2n}(K)$ and then to think of
the space of distributions as the union of all the $H^{-2n}(K)$'s. One would
then hope to \textquotedblleft dualize\textquotedblright\ all the arguments in
this paper. However, this approach requires some additional effort at each
stage, so we do not attempt to carry it out here. We hope to return to this
problem in a future paper.

\subsection{More general settings}

Finally, let us mention some additional settings in which problems similar to
those in this paper could be considered. The most natural extension would be
to consider the generalized Segal--Bargmann transform for compact symmetric
spaces, as considered in Section 11 of \cite{H1} and (in a better form) in
\cite{St}. (See also \cite{HM1}.) The necessary estimates are more complicated
in such cases; \cite{HS} is a first attempt to provide the necessary estimates.
In another direction, one could consider various Segal--Bargmann-type spaces
over $\mathbb{C}^{n},$ with various measures. Here there would be no
Segal--Bargmann transform, but one could still define holomorphic Sobolev
spaces and try to derive pointwise bounds for elements of these spaces. The
paper \cite{Si} give the first results in this direction. See also
\cite{lpbounds} for related results.

\section{Acknowledgments}

The authors thank Leonard Gross for valuable comments and suggestions.

\end{document}